\documentclass[12pt]{article}
\newcommand{\blankline}{\vskip .3cm}
\newcommand{\fa}{\begin{eqnarray}}
\newcommand{\ffa}{\end{eqnarray}}
\newcommand{\f}{\begin{equation}}
\newcommand{\ff}{\end{equation}}

\begin{document}
\centerline{\LARGE Quantum gravity with a positive cosmological 
constant}
\blankline
\blankline
\rm
\centerline{ Lee Smolin}
\blankline
\centerline{\it Perimeter Institute for Theoretical Physics}
\centerline{\it Waterloo, Canada  N2J 2W9}
\centerline{\it  \ \ and }
\centerline{\it Department of Physics, University of Waterloo}

\blankline
\blankline
\blankline
\blankline
\centerline{August 28 2002}
\blankline
\blankline
\blankline
\blankline

\centerline{ABSTRACT}
A quantum theory of gravity is described in the case of 
a positive cosmological constant in $3+1$ dimensions.  Both old and new 
results are described, which support the case that  loop
quantum gravity provides a satisfactory quantum theory of 
gravity.  These include the 
existence of a ground state, discoverd by 
Kodama, which both is an  exact solution to the constraints of quantum gravity 
and has a  semiclassical limit which is deSitter spacetime.  The long wavelength 
excitations of 
this state are studied and are shown to reproduce both gravitons and,
when matter is included, quantum field 
theory on deSitter spacetime.   Furthermore, one may derive
directly from the Wheeler-deWitt equation
corrections to the energy-momentum relations for matter fields
of the form $E^{2}= p^{2}+ m^{2} + \alpha l_{Pl}E^{3}  + \ldots$ where
$\alpha$ is a computable dimensionless constant.  This may lead in the next few 
years to experimental tests of the theory.

To study the excitations of the Kodama state
exactly requires the use of the spin network representation, which is 
quantum deformed due to the cosmological constant. The theory may be 
developed within a single horizon, and the boundary states described 
exactly in terms of a boundary Chern-Simons theory. The Bekenstein 
bound is recovered and the $N$ bound 
of Banks is given a background independent explanation.

The paper is written as an introduction to loop quantum gravity, 
requiring no prior knowledge of the subject.  
The deep relationship between quantum gravity and topological
field theory is stressed throughout.

\vfill
\blankline

\blankline
lsmolin@perimeterinstitute.ca,\\    www.perimeterinstitute.ca  
www.qgravity.org
\eject
\tableofcontents

\newpage
 
\section{Introduction}

Of the different approaches to quantum gravity, the most conservative 
is likely loop quantum gravity\cite{carlo-review,loopreviews,spain}.  
This approach is based, to begin 
with, on the quantization 
of Einstein's theory of 
general relativity, using a particular formulation discovered by Sen\cite{sen} and 
completed by Ashtekar\cite{abhay}. Rather than postulating new 
degrees of freedom, symmetries or dimensions, loop quantum gravity takes 
the basic principles 
of general relativity and quantum field theory seriously, and puts 
the emphasis on the development of methods that do not compromise 
either set of principles.  These methods highlight the force
that the principles of relativity theory, primarily diffeomorphism
invariance and the independence from any fixed, non-dynamical background
structure have, when treated properly in the context of quantum field 
theory.  Indeed, it turns out that once this is done, the theory admits
a wide range of assumptions concerning the fundamental degrees of 
freedom, symmetries and supersymmetries, as well as the exact 
dynamical laws.  Einstein's equations may be imposed, and to a 
remarkable extent, solved, 
quantum mechanically, but other assumptions concerning the fundamental 
dynamics may also be studied. 

Loop quantum gravity has been under development since 1986, and 
throughout this time there has been 
continual progress. The various obstacles encountered have in most 
cases been overcome.  As a result it has been possible over time to 
make increasingly strong claims for this approach to quantum gravity.

This paper is concerned with a subset of the results, those 
relevant for the case that the cosmological constant is non-zero
and positive.  For,
in this particular case, there 
are now, as I will describe below, 
sufficient results to claim that the theory fully deserves the 
status of a candidate for the theory of quantum spacetime.

Some of the key results on which this claim is based are old, others 
are more recent.  
The claim that loop quantum gravity with $\Lambda >0$ provides a 
consistent quantum theory of gravity includes the following:

{\bf Results special for $\Lambda >0$.}

\begin{itemize}

    \item{}The existence of an exact physical state, which exactly 
    solves all the quantum constraint equations that define the 
    theory, which also has a semiclassical interpretation, as a $WKB$
    state for deSitter spacetime. This is the Kodama state\cite{kodama}, 
    which is 
    described in section 7 below.  The existence of this state
    solves a major problem faced by loop quantum gravity,
   as it shows that, at least for $\Lambda \neq 0$, 
   the theory does have a good low energy limit which
   reproduces general relativity and quantum field theory.   
    
    \item{}Small, long wavelength perturbations of the Kodama state
    do, for small
    \f
    \lambda= \Lambda G \hbar,
    \ff
    reproduce the spectrum of gravitons on a background of deSitter 
    spacetime.    (Section 9)
    
    \item{}The theory may be coupled to arbitrary matter fields. For 
    the same conditions, long wavelength and small $\lambda$,
    perturbations of the Kodama state in the matter sector reproduce 
    $QFT$ on the deSitter background\cite{chopinlee}.  
    (Section 8) 
    
    \item{}When one extends the approximation to higher order
    terms in $l_{Pl}E$ one finds corrections to the energy-momentum
    relations for matter fields, of a form,
     \f
    E^{2}= p^{2} + m^{2} + \alpha E^{3} +\ldots
    \ff
    Such corrections are in fact amenable to
    experimental test\cite{GAC1,AC-Piron,seth,testreviews} in present and near future 
    experiments (section 10).  
    
    \item{}There is a natural boundary term\cite{linking} 
    which may be added to 
    study the case of horizons\cite{kirill1,isolated} or timelike 
    boundaries\cite{me-holo,yime-holo}. This leads to 
   an explicit construction of  the boundary Hilbert 
    space. One consequence is that the Bekenstein bound is 
    satisfied automatically\cite{linking}.  Another is a new understanding, in
    purely background independent terms, of the $N$ bound
    conjectured by Banks\cite{N-banks,N-bousso}.

\end{itemize}    

To these may be added:

\blankline

{\bf Results of loop quantum gravity for all $\Lambda$}

\begin{itemize}

    \item{}There is a detailed Hamiltonian quantization including 
    states, inner product, observables, regularization procedures 
    etc\cite{carlo-review,loopreviews,tedlee,loop1,loop2}. 
    The space of diffeomorphism invariant states is known 
    precisely and is characterized in terms of spin 
    networks\cite{sn-roger}, which 
    provide an exact orthonormal basis\cite{sn1}.  
    
    \item{}Among the operators which are understood, and which are 
    finite after a regularization procedure\cite{spain,volume}, 
    are the area of surfaces 
    (such as the boundary, when one is present) and the volume of 
    regions, including the volume of the universe. These have 
    known, discrete spectra, leading to a picture of discrete spatial 
    geometry\cite{renate-volume,volume2}.   
    Another operator which is finite and well defined is 
    the hamiltonian constraint, as well as the hamiltonian in certain 
    fixed gauges or boundary conditions\cite{ham,roumen-ham,thomas-ham}.

    \item{}A path integral formulation is known, called spin foams, 
    for computing amplitudes for evolution of spin 
    network 
    states\cite{mike-foam,RR-foam,BC,baezfoam,F-foam,tubes,foamreviews}. 
    This may be derived in several different ways, whose agreement 
    stands as evidence for the robustness of the formulation.  
    The relation to 
    the Hamiltonian theory is also understood, in the conventional 
    fashion\cite{RR-foam}.
    
      \item{}These results explain the failure of perturbative 
    approaches to quantum general relativity and supergravity, as 
    those assume the existence of states of the full theory that 
    correspond to graviton states in the linearized theory of 
    arbitrarily short wavelength. Such states do not exist in loop 
    quantum gravity, and it is fully understood why, and why this is 
    not an obstacle to the theory being consistent.

    \item{}There is a general theory of boundaries, which includes 
    horizons\cite{linking,kirill1,isolated}.  
    This leads to a complete description of black hole and 
    cosmological horizons, reproduces the Bekenstein bound and gives 
    an explanation of the Bekenstein-Hawking entropy in terms of 
    microstates per microstate, where the macrostate is the 
    classical description of an horizon and the microstates are the 
    exact states of the quantum geometry description coming from loop 
    quantum gravity. It is also possible to deduce from this 
    corrections to the Hawking radiation\cite{kirill-radiate}, 
    in particular, both 
    logarithmic corrections to the entropy formula\cite{logcorrect} and a fine 
    structure to the hawking radiation\cite{fineBH,kirill-radiate}. 
    
    \item{}There is a general approach to cosmological models, in 
    which the full quantum theory is reduced to homogeneous 
    spacetimes, called {\it loop quantum cosmology}\cite{LQC}. The results 
    reproduce those of semiclassical cosmology for times large in 
    Planck units. However the cosmological singularity is removed, and 
    a bounce is predicted\cite{LQC}.  
    
    \item{}Many of these results are confirmed by rigorous results 
    and theorems, at the level of mathematical quantum field 
    theory\cite{gangof5,thomas,thomas-ham,thomas-thesis}. 
    These explain and confirm how, from a mathematical 
    theory point of view, it is possible to find exact results about 
    states, operators, inner products etc in a diffeomorphism 
    invariant quantum field theory. 
    
    \item{}There are many checks of loop quantum gravity, made by 
    reducing the theory to special cases such as $2+1$ 
    dimensions\cite{2+1}, 
    $1+1$\cite{1+1} dimensions, the linearized theory\cite{gravitons}, 
    with and without matter 
    fields\cite{matter} etc. All 
    these checks indicate that the methods and main results are 
    reliable.  The method may also 
    be applied to Yang-Mills theory, and gives a formulation of 
    lattice gauge theory equivalent\cite{latticeloop} to the usual formulations. 
    Finally, the method of loop quantum gravity can be applied to 
    topological field theories, and reproduces results arrived at 
    by alternative methods\cite{2+1,BF}.  
    
    \item{}These methods may be applied to a wide range of theories.
    Many results extend to supergravity at least up till 
    $N=2$\cite{super,yime-holo,yi-super} and some results extend also to higher 
    dimensional theories\cite{higher,11d}. 
    All known kinds 
    of matter fields may be included\cite{matter}. There are of course special 
    results for special cases, including $3+1$ dimensions, $2+1$ 
    dimensions etc. Thus, loop quantum gravity provides a general 
    framework for formulating background independent quantum theories 
    of spacetime, gravity and matter. 
    
    \item{}This last observation leads to a strategy for finding a 
    version of loop quantum gravity that can serve as the background 
    independent formulation of string or $\cal M$ theory. This is 
    based on an observation that, as in any gauge theory, the small 
    excitations of spin foam histories around a semiclassical history 
    will be described in terms of the embedding of a string in the 
    spacetime\cite{stringsfrom}.  There are 
    definite proposals and results for this theory, based on a class 
    of matrix models, closely connected to matrix Chern-Simons 
    theory\cite{MCS}. 
    Several results exist which suggest that both perturbative string 
    theory and a form of loop quantum gravity may be derived from 
    such a  matrix model, thus realizing the conjecture of 
    duality between string and gauge descriptions also for quantum 
    gravity. 
    
    \item{}A general approach to the problems of giving a measurement 
    theory for quantum theories of cosmology, where the observer is 
    part of the system has been formulated in papers of 
    Crane\cite{louis-holo}, Rovelli\cite{carlo-relational} 
    and Markopoulou\cite{Fotini-QCH}. 
    This approach, called {\it relational quantum 
    cosmology} naturally 
    incorporates a version of the holographic 
    principle\cite{weakholo,weakstrong}.
    
\end{itemize}    

It must be said that some theoretical physicists find these results
surprising.  It used to be argued that the perturbative
non-renormalizability of general relativity implies that quantum
gravity requires a modification of the principles of physics
such as proposed in different ways in other approaches such
as string theory, causal sets, etc. 
What is perhaps surprising is that loop quantum gravity has been
successful, not by being radical, but by sticking
rather strictly to the basic principles of general relativity and
quantum theory.  

Thus, the first question to be addressed is how
it is that such results are possible, when the theory is nonsense
when developed by traditional perturbative methods around fixed 
backgrounds? The answer has two parts. First the theory is completely 
{\it background independent} which means that classical spacetimes 
play no role in the formulation. This is necessary due to the strongly
interacting nature of the Planck scale physics, the fact that the
spacetime geometry is represented completely by operators, and the fact that 
the gauge invariances of the theory include active diffeomorphisms,
which are broken by the specification of any given classical 
background metric\footnote{For background on background independence,
see \cite{dirac,stachel-bi,threeroads}.}.

The mistake all background dependent approaches make is to assume
that space and spacetime have continuous, classical structures when
probed at arbitrarily short wavelengths.  Even rough, heuristic
arguments, such as those that Wheeler and others used to give
based on the uncertainty principle, suggest that this is wrong.
The results of loop quantum gravity, based only on the basic
principles of quantum theory and general relativity demonstrate
conclusively that there is no classical spacetime manifold
at Planck scales and shorter\footnote{Indeed, the dynamics of
Einstein's equations do not come into the derivation of the
quantization of area and volume, which tell us that quantum
geometry is discrete. These results are consequences only of the
canonical commutation relations  and the gauge
invariances that define the theory. They apply whatever matter 
the theory is coupled to, and, with appropriate modifications, 
described in \cite{yime-holo} apply also to supergravity.}. 

Thus, to arrive at a good quantum theory one must use methods which are
background independent, and do not make the incorrect assumption
that spacetime is smooth at short distances.  To do this means only
to take the  basic principles of Einstein's
theory of general relativity seriously and apply them  exactly in the quantum 
theory. As opposed to other approaches, such as string theory,
where new degrees of freedom are posited to move in fixed, classical
background spacetimes, loop quantum gravity treats the geometry of
spacetime at the quantum mechanical level as Einstein did at the 
classical level, as a completely dynamical entity. This is in fact
required if the gauge invariance of the theory, which is active
diffeomorphisms, is to be respected exactly in the quantum theory.

When diffeomorphism 
invariance is imposed exactly there is a big payoff, which is that one 
finds an exact description of the gauge invariant Hilbert space of 
quantum gravity. These are eigenstates of observables that represent 
the volume of spacetime and the areas of surfaces in the spacetime. 
These observables turn out to be finite, when regulated in a manner 
that respects diffeomorphism invariance. And they have discrete 
spectra, which demonstrates that quantum geometry is discrete at the 
Planck scale. These results explain, in detail, why perturbative 
expansions around smooth backgrounds fail; because they fail to 
capture any of the structure present in exact solutions of the quantum
constraints that impose gauge and diffeomorphism invariance.  

The second key to the success of loop quantum gravity is a property
of the Einstein equations, which gives force to these general considerations 
of background independence and diffeomorphism invariance. This 
is the existence of an intimate connection between the kinematics and 
dynamics of general relativity and topological field theories. 

A topological field theory is a field  theory that has only a finite
number of degrees of freedom. The few degrees of freedom it has
are non-local and generally
are measured either at boundaries of spacetime or by measuring
phase factors or holonomies associated with loops or surfaces
that cannot be shrunk to a point.   Topological field theories
share some properties with general relativity, such as being
invariant under the gauge group of active diffeomorphisms, and
being independent of any classical background\footnote{Some basic 
references on topological field theory include\cite{basic-tft}}.

At the same time,  general relativity is not a topological field 
theory as it has an infinite number of local degrees of 
freedom\footnote{There are other diffeomorphism invariant theories
that are not topological, even some without a metric, such as 
Chern-Simons theory for $d \geq 5$\cite{5cs} and higher form versions
of Chern-Simons theory\cite{highercs,11d}.}
Even so, there are close connections between general relativity
and topological field theory at the classical level, and these
are exploited by loop quantum gravity to make a sensible
quantization of general relativity. By doing so, 
results that at first sight are surprising and unexpected, turn out to
be not only possible, but accessible with standard methods. 
This  relationship with topological field theory
in fact makes possible the key results at both the hamiltonian and path 
integral level that are the basis for the success of the approach.

These relationships with topological field theory are the subject of 
the first few sections of this paper, as they are in a good way
into the subject. There are in fact three distinct ways that 
topological field theory enters into quantum gravity, First, 
the action turns out to be closely related to that of a topological 
field theory\cite{spain}. 
Second, the natural boundary term in the action, which 
must be added when a spacetime region with 
boundary is studied, is a topological field theory\cite{linking}. Third,
the ground state of the theory with a non-zero $\Lambda$ 
is derived from a topological field theory\cite{kodama}. 
 
Only a few of these results are recent.
The case of positive cosmological constant has, in 
fact, been somewhat neglected in the field in recent years.  Thus it 
is worth mentioning some of the 
reasons to return to it now. These include the fact that a positive
cosmological constant appears to be observed
(not too mention the fact that it is in 
any case necessary for inflationary cosmology that the effective 
cosmological constant at early times be positive, and large).  Beyond this, 
general renormalization group arguments tell us that we cannot make sense of 
any quantum field theory unless we include all low dimension couplings.  This is 
certainly the case with the cosmological constant; if it is not 
included in a generic quantum field theory it will be there in the 
effective theory with a magnitude order one in either the cutoff 
or the supersymmetry breaking scale. This general expectation is 
fully born out in all the 
numerical work done on nonperturbative approaches to quantum 
gravity. This includes the early work in euclidean dynamical 
triangulations as well as the recent numerical work of Ambjorn, Loll 
and collaborators on lorentzian, or causal, dynamical 
triangulation models\cite{AL}.  Further the 
latter calculations show convergence 
is only possible if $\Lambda >0$.  Indeed, all the evidence we have from
explicit renormalization group calculations, at both the perturbative
and non-perturbative levels, tells us that the theory cannot have
a good low energy limit, leading to the recovery of general 
relativity, unless the cosmological constant term is included in the
formulation of the theory. 

From the point of view of loop quantum gravity, the main difference
so far discovered between the theory with and without the cosmological
constant is that it is only when $\Lambda \neq 0$ that the Kodama
state exists. This is the only known exact solution of the theory that
also is known to have a well behaved low energy description, in terms
of semiclassical physics. For the case $\Lambda =0$ the demonstration
of the existence of a good low energy limit remains an open 
problem, although  progress has been achieved recently
on related technical issues such as the formulation of the
renormalization group in spin foam models\cite{fotini-rg} and
the construction of coherent states of the quantum gravitational
field\cite{thomas-thesis}.

Closely connected with this issue has been a question of whether
there could be long ranged correlations in the dynamics generated
by the Hamiltonian constraint\cite{problem1,problem2}. 
This problem is solved for the
case of $\Lambda \neq 0$, in a rather elegant fashion that will
be described in section 12 \cite{tubes}.  

As the demonstration of a good low energy limit has been a key open
problem, it is of some importance
that the problem is solved when $\Lambda \neq 0$.  It is because
of this that it can now be claimed that loop quantum gravity
passes all the major tests required of a good quantum theory of
gravity\footnote{For a comparative evaluation of the results of
the different approaches to quantum gravity, see \cite{mewheeler}.}.  

If further reasons after this are needed, we can mention the fact that 
string theory appears to have trouble with this case, so it may be one 
in which there is no alternative to background independent methods. 
Finally, this is a good case to study the problem of interpretational issues 
for quantum cosmology and illustrate how {\it relational quantum cosmology} 
addresses them.
  
This paper is written with the hope that it may serve as an 
introduction to this field, and for this reason the style is
pedagogical and many old
results are included.  I assume that the reader has no prior knowledge
of loop quantum gravity,  and knows only the basics of
Yang-Mills theory, general relativity and quantum theory.  
Many technical details are omitted, and may be found in the 
references indicated.  The goal is 
instead to give the reader clear statements of the assumptions and 
results, sufficient to serve as an introduction to the field. 

However, the paper is not completely a review. 
There are included a few new results, 
which support earlier claims
that the theory is well defined at both the exact and semiclassical 
level.  The new results are in sections 3, 9,10 and 13 and are also
discussed in papers in preparation. 

The paper has two parts, the first describes the background
of classical general relativity, stressing the relationship to
topological field theory. The second part describes the quantum
theory of gravity with $\Lambda >0$.  The table of contents may be 
consulted for a more detailed description of the contents.

\part{Classical gravity and its relation to topological field theory}

\section{Gravity as a gauge theory}

Let us jump right in and see the power of the connection between
gauge theory, gravity and topological field theory uncovered
in loop quantum gravity, and then go on to show how this perspective
illuminates the geometry of deSitter spacetime. 

A good way into the subject is to begin with the following challenge:
{\it Suppose that you wanted to make a theory of gravity, but you were
restricted to using only the fields of an ordinary gauge theory.  
You are not allowed to assume the existence of a metric, either as a 
background or as a dynamical field. You have to work only with a gauge field.
How close would you come to general relativity?}

The answer is that the simplest guess as to how to do this lands
right on the nose on general relativity. Here is how this goes:

It turns out to be most direct
to reason in the Hamiltonian language. From this point of view a 
spacetime is a manifold of the form $\Sigma \times R$, where $\Sigma$
is a three dimensional manifold which will represent space, at least
topologically.  

From a hamiltonian perspective the fields we are allowed are a
connection, $A_a^i$, and its conjugate momentum,
$E^a_i$ where $a$ is a $3d$ spatial index and $i$ is valued in a lie
algebra, $G$.

Thus, we have the Poisson brackets,
\f
\{ A_a^i (x) , E^b_j (y) \} = \delta_a^b \delta^i_j \delta^3 (y,x)
\label{ccr}
\ff

We know that in a hamiltonian formulation of a gauge theory there is 
one constraint for each independent gauge transform\cite{dirac}.  The
gauge invariances of a gravitational theory include at least
$4$ diffeomorphisms, per point.  Thus,

\f
I^{GR}= \int dt \int_\Sigma  E^{ai} \dot{A}_{ai} - N{\cal H} -N^a 
{H}_a - w_i {\cal G}^i - h
\ff
where ${\cal H}_a$ generates the diffeomorphisms of $\Sigma$, $\cal H$ 
must be the so-called Hamiltonian constraint that
generates the rest of the diffeomorphism group of the spacetime (and 
hence changes of the slicings of the spacetime into spatial slices) 
while
${\cal G}^i$ generates the local gauge transformations. $h$ 
represents the terms in the hamiltonian that are not
proportional to constraints.  However, there is a special feature of
gravitational theories, which is there is no way {\it locally} to
distinguish the changes in the local fields under evolution from
their changes under a diffeomorphism that changes the time coordinate.
Hence $h$ is always just a boundary term, in a theory of gravity.

From Yang-Mills theory we know that the constraint that
generates local gauge transformations under (\ref{ccr}) is
just Gauss's law
\f
{\cal G}^i = {\cal D}_a E^{ai} =0
\label{gausslaw}
\ff
Note that $E^a_i$ is a vector {\it density}, so there is no metric
used in either the Poisson brackets or Gauss's 
law\footnote{One thing to get used to in this field is that as 
there is no background metric, while in the quantum theory the
metric is a composite
operator, one must be completely explicit about
all places the metric appears and all density weights.}. 

Let us now guess the forms of the other constraints. First there must
be three constraints per point that generate the diffeomorphisms of
the spatial slice. Infinitesimally these will look like coordinate
transformations, hence the parameter that gives the infinitesimal
change is a vector field. Hence these constraints must multiply a vector
field, without using a metric. Thus these constraints are the
components of  a one form. It should also be invariant
under ordinary gauge transformations, as they commute with
diffeomorphisms. We can then ask what is the simplest
such beast we can make using $A_a^i$ and $E^a_i$? The
answer is obvious, it is
\f
{\cal H}_a = E^b_{i} F_{ab}^{i} =0
\label{diffeoconstraint}
\ff
where $F_{ab}^i$ is
the Yang-Mills field strength.

It is a simple exercise to show that ${\cal H}_a$ so defined does
in fact
generate a spatial diffeomorphism (plus an ordinary gauge
transformation) on the fields $A_a$ and $E^a$.

There remains one constraint per point, which generates changes in the
time coordinate, or else in the embedding of $\Sigma$ in 
${\cal M} = \Sigma \times R$. This is called the Hamiltonian
constraint. Since its action is locally indistinguishable from
the effect of changing the time coordinate, it does contain
the dynamics. 

The Hamiltonian constraint must be gauge invariant and a scalar, since
the parameter it multiplies is proportional to the local change
in the time coordinate.  But it could also be a density, so we have
the freedom to find the simplest expression that is a density of some
weight. It turns out there are no polynomials in our fields that
have density weight zero, without using a metric. But there are
simple expressions that have density weight two. 
The two simplest such terms that can
be written, which are lowest order in derivatives, are, 
\f
\epsilon_{abc}TrE^aE^bE^c  \ \ \ \mbox{and} \ \ \ Tr E^a E^b F_{ab}
\ff
where the $Tr$ is in the lie algebra $G$.  
If we need to we could go to terms with more derivatives, but such
terms will give trouble if we want the theory to have a simple 
linearization, which will be useful to reproduce Newtonian gravity
and gravitational waves.  

In fact these two terms already
give Einstein's equations, so long as we take the simplest
nontrivial choice for $G$, which is $SU(2)$.

Thus, we take for the Hamiltonian constraint
\f
{\cal H} = \epsilon_{ijk} E^{ai}E^{bj} 
(  F^k_{ab} + {\Lambda \over 3} \epsilon_{abc} E^{ck} ) =0
\label{hamconstraint}
\ff

There is a place to put a free parameter $\Lambda$ which indeed
will turn out to be the cosmological constant. As far as dimensions 
are concerned, $A_a$ is a connection and so has dimensions of
inverse length. It will turn out that $E^a$ is related to the metric
and so we should make the unconventional choice that it is 
dimensionsless. 

In fact, what we have here is Euclidean general relativity.
If we want the Lorentzian theory, we need only modify what
we have by putting an $\imath$ into the commutation relations,
so we have instead of (\ref{ccr})

\f
\{ A_a^i (x) , E^b_j (y) \} = \imath G \delta_a^b \delta^i_j \delta^3 (y,x)
\label{lccr}
\ff
I have also inserted a factor of Newton's constant, $G$, which is
necessary if $E^a$ is dimensionless.

The Einstein's equations of course come from taking Poisson
brackets with the Hamiltonian, which is a linear combination of
constraints,
\f
H(N,v^a) = \int_\Sigma N {\cal H} + v^a {\cal H}_a
\ff
here $N$ and $v^a$ are related to the usual lapse and shift,
which are in turn related to the time-time and time-space components
of the metric, respectively.  In fact, noting that $\cal H$ has
density weight two, we see that $N$ must have density weight $-1$.
Hence it is of the form of $g_{00}/\sqrt{detq_{ij}}$, where
$q_{ij}$ is the spatial metric. (This may seem pedantic but we will
use it to good effect in the next section.)

The simplest way to evolve is with zero shift, which corresponds
to the space-time components of the metric vanishing. The equations 
of motion are then,
\f
\dot{A}_{ai} = \{ A_{ai} , \int N {\cal H} \}  =N\imath G  \epsilon_{ijk} E^{bj} 
( 2 F^k_{ab} + {\Lambda } \epsilon_{abc} E^{ck} )
\label{Adot}
\ff
\f
\dot{E}^{ai} = \{ E^{ai} , \int N {\cal H} \} = \imath G 
\epsilon^{ijk} {\cal D}_b (N E^a_j E^b_k )
\label{Edot}
\ff

These equations, together with the seven constraints make a
diffeomorphism invariant field theory whose only degrees of 
freedom are an $SU(2)$ connection and its conjugate electric
field. To see that the theory is consistent one should check
the constraint algebra, in fact it is first class. One can then
count degrees of freedom and find that there are $2$ canonical
degrees of freedom per point.  If one linearizes, one gets right
away the laws for the propagation of
spin $2$ massless fields. 

How can this be, when there is no metric in the world our equations
describe? In fact there is one, it is hidden in the gauge fields.
The theory we have is general relativity, with the following
identifications.  $E^a_i$ is related to the three metric $q_{ab}$
by
\f
det(q)q^{ab}= E^{a i} E^{bj} \delta_{ij}
\ff
The determinant is there because the expression is a density of weight two. 

The $SU(2)$ connection $A_a$ turns out to be, for solutions,
the self-dual part of the spacetime connection. For Lorentzian
solutions this is complex, and its real and imaginary parts each
have a geometrical interpretation. 
\f
A_{ai}= \mbox{3d spin connection}_{ai}
+ {\imath  \over \sqrt{q}} K_{ab}E^b_i
\label{ashtekar}
\ff
where $K_{ab}$ is the extrinsic curvature of the $3$ manifold $\Sigma$
embedded in the spacetime, which in turn is essentially the time
derivative of the three metric\footnote{For more details on the
canonical formulation of GR in Ashtekar-Sen variables, see
\cite{sen,abhay} as well as the books\cite{abhay-books}.}.

\section{The deSitter solution as a gauge field}

It is not of course obvious to see that the theory we have constructed
is in fact  Einstein's theory, or where the correspondences I've just 
mentioned come from.  A bit later we will derive these from an 
action principle. But for now I want to only show how the deSitter
solution fits into this framework.

We begin by noting that a family of solutions to the
constraints can be read off immediately, by inspection. These are those
that satisfy, 
\f
F^i_{ab} =   -{\Lambda \over 3} 
\epsilon_{abc} E^{ci}
\label{selfdual1}
\ff
It is easy to see that they satisfy all seven\footnote{$3$ generate 
spatial diffeo's, three generate $SU(2)$ gauge transformations plus
the Hamiltonian constraint.} constraints. We call 
these {\it self-dual solutions} as they have the magnetic fields
proportional to the electric fields. 

Let us examine the simplest one of these. We can take as an ansatz
that $A_a^i$ is proportional to $\delta_a^i$. Of course this breaks 
gauge invariance but this is what we have to do if we want to write
an explicit solution. 

As I know the answer, I will put in the right parameters: 

\f
A_{ai}= \imath \sqrt{\Lambda \over 3} \  f(t) \  \delta_{ai} \ \ \mapsto 
F_{abi} = -f^2 (t) \ {\Lambda \over 3} \epsilon_{abi} 
\label{Abackground}
\ff
where $f(t)$ is a function of the time coordinates.  

Taking $A_a^i$ to be purely imaginary makes sense in light of
(\ref{ashtekar}), it means that we are making an ansatz that the
three geometry is flat, so the three dimensional spin connection
vanishes. The metric can then be taken to be homogeneous, as must
also be its time derivative, which is the imaginary part of $A_a^i$.

By the self-dual initial conditions we see that 
\f
 E^{ai}=  f^2 
\delta^{ai} \ \ \ \mapsto \ \ \  q_{ab} = f^2 \delta_{ab}
\label{Ebackground}
\ff
 
As we have satisfied the self-dual condition all the constraints
are satisfied. We merely have to plug into the equations of motion
(\ref{Adot},\ref{Edot}) to find the evolution equations for $f$.  
Both equations
of motion agree that 
\f
\dot{f}= N \sqrt{\Lambda \over 3 } f^4  
\ff
Remembering that $N$ is an inverse density, we should take
$\mapsto N \approx  det(q)^{-1/2}  = f^{-3}$. This gives us
\f
\dot{f}=N   \sqrt{\Lambda \over 3} f^4 =  \sqrt{\Lambda \over 3} f 
\ff
so that $f=e^{ \sqrt{\Lambda \over 3}t}$.
  
With the identifications we have made this 
gives the deSitter spacetime in spatially flat 
coordinates\footnote{A good review of the different coordinatizations
of deSitter spacetime is in \cite{AndyetdS}.}: 
\f 
ds^2 = -dt^2 + e^{2 \sqrt{\Lambda \over 3} \ t}(dx^a)^2 
\ff

\section{Hamilton-Jacobi theory, deSitter spacetime and Chern-Simons
theory}

Before we get serious and go back to the action and show why this
is really Einstein's theory, there is one more simple trick we can
do, which brings to light a connection between deSitter spacetime and 
topological field theory. 

To see this connection we may begin by 
asking what insight Hamilton-Jacobi theory may
throw on the solutions we have been considering. To use Hamilton-Jacobi
theory we assume that there is a Hamilton-Jacobi functional $S(A)$
on the configuration space. As we are studying a gauge theory the
configuration space is the space of the connections $A_a$ on the
three manifold $\Sigma$. 

The conjugate electric field must then be the gradient of the
Hamiltonian-Jacobi function, 
\f
 E^{ai} = -{\delta S(A)  \over \delta A_{ai}}
\ff

We found that all seven constraints are solved with the self-dual
ansatz (\ref{selfdual1}).  This means that the Hamilton-Jacobi function must
satisfy a first order differential equation,
\f
F^i_{ab} = - {\Lambda \over 3} \epsilon_{abc} E^{ci} 
= {\Lambda \over 3} \epsilon_{abc} {\delta S(A)  \over \delta A_{ci}}
\ff

This integrates immediately to
\f
S_{CS}= {2\over 3 \Lambda} \int Y_{CS}
\ff
Here $Y_{CS}$ is the famous 
Chern-Simons invariant, given by  
\f
Y_{CS} = {1\over 2} Tr ( A\wedge dA + {2\over 3} 
A^3 ).
\ff

It satisfies $ {\delta \int Y_{CS}\over  \delta A_{ai}} = 
2 \epsilon^{abc}F_{bc}^i $

Thus, {\it the self-dual solutions follow trajectories in configuration
space which are gradients of the Chern-Simons invariant.} Not only
is deSitter spacetime one of these, there is the remarkable fact that,
while there are an infinite number of self-dual solutions for
Euclidean signature, there is only one for Lorentzian signature
and it is deSitter spacetime.

This suggests that the semiclassical state that describes deSitter is
\f
\Psi_{K} (A) = {\cal N} e^{{3 \over 2 \lambda} \int Y_{CS}}
\ff

${\cal N}$ is a normalization depending only on 
topology\cite{chopin-new}.

In fact, {\it this is an exact quantum state} as was shown
in 1990 by Hideo Kodama\cite{kodama}.  
We will return to the Kodama state and the physics that may be
derived from it. But first we want to go back and find out why
what we have been studying is Einstein's general theory of 
relativity.

\section{General Relativity as a constrained topological 
    field theory}
    
    In the last sections a very mysterious fact emerged, which is that
    when general relativity is written in such a way as to bring it
    close to gauge theory, in terms of field content and geometry,
    we fell upon a close relationship between an important set of 
    solutions-the self-dual solutions, and a topological field theory.
    Given the ease by which topological field theories may be 
    quantized and studied, as well as their remarkable connections
    with various fields of algebra, representation theory and topology,
    it is very important to know if this is an accident or if it
    has its roots in some deep relationship between gravity and
    topological field theory. In this section we will show that it is 
    indeed no accident and that general relativity and topological field 
    theory are deeply connected at the level of the action principle.

   \subsection*{$BF$ theory}
   
   We begin with a four dimensional topological field 
   theory called $BF$ theory\cite{BF}.
   We will work on a four manifold ${\cal M}= \Sigma \times R$,
   where $\Sigma$ will be the spatial topology.  There is no metric,
   and no other fixed background field.
   
   We introduce now two fields. The first is an 
   $SU(2)$ connection $A_{\mu}^{i}$, where $\mu$ indices the spatial
   coordinates (to be suppressed when we use form notation and
   $i=1,2,3$ label the generators of $SU(2)$.  The second field
   is a two form, $B^{i}_{\mu \nu}$ which is also valued in the
   $SU(2)$ generators.  To begin with we take them both real. 
    
The action we use is,   
    
\f
I^{BF}= \int B^{i}\wedge F_{i} + {\Lambda \over 2}  B^{i}\wedge B_{i}.
\label{BF}
\ff

It is easy to derive the equations of motion, 
\f
F^{i} = -\Lambda B^{i}  , \ \ \ \ \ \ \ \ 
{\cal D} \wedge 
B^{i}=0
\label{bfeom}
\ff
We see that the curvature is constrained to be proportional to the
$B$ field, with $\Lambda$ the constant of proportionality.
$B^{i}$ is in turn is constrained to be covariantly constant.
If one counts one finds there are no local degrees of freedom, hence
the theory is topological.  It is also invariant under 
$Diff({\cal M})$, the group of diffeomorphisms of the 
manifold.  
Because of the form of the action, this topological field theory
is called $BF$ theory.

\subsection*{Self-dualology}

General relativity is in fact closely related to $BF$ theory. To
see this, we need first to understand the dynamics of general
relativity in $4$ spacetime dimensions 
in terms of self-dual and antiself-dual connections and 
curvatures.  

Let us then have a four dimensional spacetime metric
$g_{\mu \nu}$. We will at first take the spacetime to be Euclidean,
then we will see how things are modified for the Lorentzian case.

It is convenient to work with frame fields, 
$e^{a}_{\mu}$, with $a=0,1,2,3=0,i$ being four dimensional
frame field indices.  
They are related to the metric by
\f
g_{\mu \nu}= e^{a}_{\mu}e^{b}_{\nu} \eta_{ab}
\ff
with $\eta_{ab}$ the flat metric on the tangent space.

Now we need do do a little {\it  self-dualology.} Let us consider an 
antisymmetric tensor $A_{ab}$ in the tangent space. Given the
totally antisymmetric $\epsilon^{abcd}$ and the metric $\eta^{ab}$
we may divide $A_{ab}$ into its self-dual and antiself-dual parts
\f
A_{ab}^{\pm}= {1\over 2} \left (
A_{ab} \pm A^{*}_{ab}
\right )
\ff
where $A^{*}_{ab}= {1\over 2} \epsilon_{ab}^{ \ \ cd} A_{cd}$.
We have
\f
(A_{ab}^{\pm})^{*}= \pm A_{ab}^{\pm}
\ff
Note that these equations are consistent with $**=+1$, which is
the case for Euclidean signature.

Among the objects that can be decomposed this way are the
spin connection one form $A_{ab}$ and the curvature
two form 
$F_{ab}= dA_{ab} + {1\over 2} A_{a}^{\ c} A_{bc}$. These are
valued in the $SO(4)$ lie algebra. 
The decomposition of $A_{ab}$ into $A_{ab}^{+}$ and $A_{ab}^{-}$
corresponds to the Lie algebra identity  $SO(4)=SO(3)_{L} \oplus 
SO(3)_{R}$. There are then three generators (per form index)
in $A_{ab}^{+}$ and they correspond to $SO(3)_{L}$. These three
generators may then be labeled by $i=1,2,3$ by the correspondence
$A_{i}^{+}=A_{0i}^{+}= {1\over 2} \epsilon_{i}^{\ jk} A_{jk}^{+}$.

It is important to note that $F^{+}_{i}$, being also valued
in $SO(3)_{L}$ is an $SO(3)$ gauge field which is a function only
of the $SO(3)_{L}$ connection $A^{+}_{i}$.  

It turns out that not only can the connection and curvature 
information in a four manifold be decomposed in self-dual and
antiself-dual parts, the same is true for the metric information. 
Given the metric $g_{\mu\nu}$ one can construct three
two forms from the self-dual parts of $e^{a} \wedge e^{b}$, as
\f
\Sigma^{i} = e^{0}\wedge e^{i} + \epsilon^{i}_{\ jk} e^{j}\wedge e^{k}
\ff
 
These forms are self-dual by construction in the internal indices.
Each of the three is also {\it self-dual in the spacetime sense}
\f
{}^{*}\Sigma_{\mu \nu}^{i} \equiv \epsilon_{\mu \nu \lambda \sigma}
g^{\lambda \alpha}g^{\sigma \beta} \Sigma_{\mu \nu}^{i}=
\Sigma_{\alpha \beta}^{i}
\ff
 
\subsection*{From self-dual two forms to general relativity}

The connection of general relativity to $BF$ theory comes about
by identifying the $SO(3)$ valued $B^{i}$ fields, which are
three two forms, with the self-dual two forms $\Sigma^{i}$ 
corresponding to some metric $g_{\mu\nu}$. Let us see
how this works.

To make the correspondence we cannot just plug
\f
B^{i}= \Sigma^{i}= 
e^{0}\wedge e^{i} + \epsilon^{i}_{\ jk} e^{j}\wedge e^{k}
\label{correspond}
\ff
into the equations of motion, (\ref{bfeom}),
as the restriction (\ref{correspond})
reduces the number of degrees of freedom per point, and there are 
already zero degrees of freedom per point. But we can plug
(\ref{correspond}) into the action (\ref{BF}) for $BF$ theory to find
that,
\f
I^{JSS}= I^{BF}|_{B^{i}= \Sigma^{i}}= 
\int ( e^{0}\wedge e^{i} + \epsilon^{i}_{\ jk} e^{j}\wedge 
e^{k} )  \wedge F_{i} + {\Lambda \over 2} 
\epsilon_{abcd}e^{a} \wedge e^{b} \wedge e^{c} \wedge e^{d}
\label{JSS}
\ff
This is actually an action for general relativity\cite{JSS}. 
In fact it is easy 
to see that it gives the same equations of motion as the Palatini
action
\f
I^{Palatini}=   
\int \epsilon_{abcd} \left (  e^{a} \wedge e^{b}  \wedge F^{cd} 
+ {\Lambda \over 2} 
e^{a} \wedge e^{b} \wedge e^{c} \wedge e^{d} \right ) 
\label{palatini}
\ff

Using the 
projections into the self-dual and antiself-dual parts of the
curvature, our strange looking action (\ref{JSS}) can be written as, 
\f
I^{JSS}=   
\int \epsilon_{abcd} \left (  e^{a} \wedge e^{b}  \wedge F^{+ cd} 
+ {\Lambda \over 2} 
e^{a} \wedge e^{b} \wedge e^{c} \wedge e^{d} \right ) 
\label{JSS2}
\ff
The equations of motion that come from varying the self-dual
part of the connection, $A^{+}_{i}$ are
\f
({\cal D}^{+} \Sigma )^{i}=0
\label{halfconn}
\ff
These three equations are in fact the self-dual projection of the 
six equation of motion that
corresponds  to varying the Palatini action by the 
full $SO(3) \oplus SO(3)$ connection, $A^{ab}$ to find,
\f
\nabla e^{a}\wedge e^{b} =0
\label{palconn}
\ff

It is well known that the solution to this last (\ref{palconn}) is 
that $A^{ab}$ is equal to 
the $SO(4)$ spin connection, $\omega^{ab}$ corresponding to the
frame field $e^{a}$.  The solution to the equations of motion
of the modified action (\ref{halfconn}) are similar, they are that
$A^{+}_{i}$ is equal to $\omega^{+ ab}$, which is the 
self-dual part of the spin connection of $e^{a}$.

The other equation of motion of the Palatini equation is, with
the connection taken to be the spin connection, the Einstein 
equations, 
\f
\epsilon_{abcd} \left (  e^{b}  \wedge F^{ cd} 
+ { \Lambda } 
e^{a} \wedge e^{b} \wedge e^{c} \wedge e^{d} \right ) =0 .
\label{einstein}
\ff
The equation of motion of the modified equation is instead
\f
\epsilon_{abcd} \left ( e^{b}  \wedge F^{+ cd} (\omega^{+} )
+ { \Lambda } 
e^{a} \wedge e^{b} \wedge e^{c} \wedge e^{d} \right ) =0
\ff
This differs from the Einstein equation (\ref{einstein}) by
a single term, which is 
\f
e_{c}  \wedge F^{ cd} (\omega ) , 
\ff
but this vanishes by the Bianchi identity that sets 
$R_{\mu[\nu \lambda \sigma ] } =0$. 

This establishes the equivalence of (\ref{JSS2}) to general 
relativity with Euclidean 
signature\footnote{Except that, as in the Palatini case, the fact
that the action and equations of motion are polynomial means there are
solutions when $det(e)=0$ that would not be non-degenerate solutions
of general relativity. Thus the space of solutions has been expanded 
by the addition of a kind of boundary that includes solutions with
degenerate frame fields.}.

\subsection*{Back to $BF$ theory}

We are not yet done for as discovered, first by 
Plebanski\cite{Plebanski}, and then by
Capovilla, Dell and Jacobson\cite{CDJ}, we can use what 
we have just learned to 
put the action in a form that shows a direct relationship to $BF$ 
theory. To do this we ask whether there are conditions on the 
two form fields $B^{i}$ which are sufficient for there to 
exist a metric, and hence a frame field, $e^{a}$ such that
$B^{i}$ are the self-dual two forms of $e^{a}$.  The answer is
yes, these are the five equations
\f
B^{(i} \wedge B^{j)} = {1\over 3} \delta^{ij}B^{k} \wedge B_{k}
\label{five}
\ff
This is easy to see one way, by plugging in, for the other, see 
\cite{CDJ}.  

Why five equations? There are $18$ components in the $B^{i}$'s 
minus three gauge degrees of freedom for the
$SO (3)$ rotations that mix them up, minus
five equations yields the $10$ components of the metric $g_{\mu\nu}$.

Thus, general relativity is the consequence of varying the $BF$ action
with the $B^{i}$ fields subject to the five constraints, (\ref{five}).
Thus, if we add these constraints times lagrange multipliers to the
$BF$ action, we get an action for general relativity in the form, 
\f
I^{Plebanski}= 
\int B^{i}\wedge F_{i} + {\Lambda \over 2}  B^{i}\wedge B_{i} -
{1\over 2} \phi_{ij} B^{i} \wedge B^{j}
\ff
so long as $\phi_{ij}$ itself is constrained to be symmetric
and tracefree. 

Actually we can incorporate the cosmological constant term
by requiring instead that
\f
\phi_{i}^{\ i} = - \Lambda ; \ \ \ \ \    \phi_{[ij]}=0
\label{constraint}
\ff
so that the action is now,
\f
I^{Plebanski}= 
\int B^{i}\wedge F_{i}  - {1\over 2}
\phi_{ij} B^{i} \wedge B^{j}
\label{plebanski}
\ff

Although we will not need it in what follows, it is interesting
to note that since the action is quadratic in the $B^{i}$ these
can be integrated out (or solved for) to find an even simpler
form for the action
\f
I^{CDJ}= \int F^{i} \wedge F^{j} (\phi^{-1})_{ij}
\label{CDJ}
\ff
Thus, we see that an action can be written for general relativity
with a non-zero cosmological constant, {\it in which the metric does not 
appear at all.}  All that appears is the curvature of the left handed 
part of the spin connection, and a new field $\phi_{ij}$, whose
trace is constrained by (\ref{constraint}) to be the
cosmological constant.  The metric is instead a {\it composite field},
which arises only for solutions of the equations of motion. 

And what is the physical interpretation of the new field
$\phi_{ij}$?  To answer this we need only look at the
field equation gotten from varying $B^{i}$ in
the Plebanski action, eq. 
(\ref{plebanski}). 
\f
F^{i}= \phi^{i}_{\ j} B^{i}
\ff
Since we learn by varying $\phi$ that there exists a metric,
whose self-dual two form $B^{i}$ becomes, we learn that
the $\phi_{ij}$ are just the components of the self-dual half
of the curvature two form, when expanded in components of the
frame fields, or equivalently, directly in terms of the
self-dual two forms of the metric.  So the action 
(\ref{CDJ}) codes for the metric in the backhanded way that
the $\phi_{ij}$ have to turn out to be the components of
the curvatures $F^{i}$ expanded in frame field components of
that metric. Very sneaky, but effective, as we shall see.

\subsection*{The same thing with Lorentzian signature}

So far everything was presented assuming the metric has Euclidean
signature. But for real physics we need the metric to be lorentzian.

The same steps yield a connection between lorentzian general relativity
and $BF$ theory, but it is a bit more complicated because all the 
fields become complex. To understand this it is best to proceed in two
steps. First, we go back and fix the definition of self-dual fields.
This is necessary because, as may be easily checked, for Lorentzian
signature $**=-1$. To accommodate this, we must insert an 
$\imath$ into the definition of self-dual tensors, 
\f
A_{ab}^{\pm}= {1\over 2} \left (
A_{ab} \pm \imath A^{*}_{ab}
\right )
\label{lorentzsd}
\ff
Thus, we now have
\f
(A_{ab}^{\pm})^{*}= \pm \imath A_{ab}^{\pm}
\ff

This means that the self-dual connection and curvature components
$A^{+}_{ab}$ and $F^{+}_{ab}$ are now complex.  That is, the
left handed part of an $SO(3,1)$ connection is really a complex
one form valued in the complexification of $SO(3)$.  

Due to the $\imath$ in eq. (\ref{lorentzsd}) the action now
has the form, 
\f
I^{Lorentz}= 
 \imath \int B^{i}\wedge F_{i}  - {1\over 2}
\phi_{ij} B^{i} \wedge B^{j}
\label{lorentz-pleb}
\ff

One may wonder whether the fact that $A^{+}$ and $A^{-}$ are both
complex has doubled the degrees of freedom. The answer depends on
whether or not we want the spacetime frame fields $e^{a}$ to be
real. If we don't require the frame fields to be real then we have
extended the theory to allow all solutions of Einstein's equations 
where the metric is complex. In this case we have doubled the number
of degrees of freedom. However, if we want the metric, and the frame
field components to be real than there is a restriction on the
self-dual and antiself-dual components, coming from the fact
that the spin connection of the metric is real. Thus, we have
\f
\bar{A}^{-}_{i}= A^{+}_{i}
\ff
This is an important difference from the Euclidean theory, in which
$A^{+}$ and $A^{-}$ are independent, but both real. 

Nevertheless, we can proceed as follows.  We can consider the
$JSS$ action, (\ref{JSS2}) for the case of real $e^{a}$ but complex $A^{+}_{i}$.
The equations of motion then still constrain the $A^{+}_{i}$
to be the self-dual parts of the spin connection and since the
$e^{a}$ are assumed real we still get the real lorentzian Einstein
equations from (\ref{JSS}). The only difference is there should be an
$\imath$ in front of the whole action. 

The next stage is to eliminate the metric completely from the action,
by going to the Plebanski action (\ref{plebanski}). Here there is
no simple way to incorporate the condition that the metric is
real and lorentzian.  The problem is that the
$\phi_{ij}$'s are complex for real, lorentzian metrics. 
The simplest thing to do seems to be to
simply consider the actions (\ref{plebanski}) and (\ref{CDJ}) for
complex fields $A^{i}$ and $\phi_{ij}$.  One then gets the complexified
lorentzian Einstein equations. One can then add to the field equations
the initial conditions that the frame field or metric components
are real.  The field equations are complex, but they have the property
that restricted to initial data in which the metric is real, the 
solutions will conserve the reality of the metric. 

This may seem a strange thing to do, but from the point of view of the 
quantum theory it is not so bad to split the field equations into
polynomial equations, which are complex, plus reality conditions on 
the fields. The reason is that in quantum theory the reality 
conditions become hermiticity conditions on operators, and these
are different from the operator equations of motion, in that they
involve the inner product. Strictly speaking, in quantum theory one
always works with the complexified operator algebra, and imposes
reality conditions through the choice of the inner product. So to
do the same in quantum gravity is not very much of an 
innovation\footnote{For this reason, in the early days of loop quantum 
gravity the strategy of expressing the reality conditions on the metric
only through the inner product, while working with a complex self-dual 
connection, seemed a good one as it greatly simplified the dynamics 
and led to many new results.  More recently another alternative was
adopted in many calculations, in which one worked with another
$SO(3)$ connection, which is real, invented by Barbero\cite{barbero}.  
This leads to more complicated constraint equations which, however,
Thiemann showed were still manageable\cite{thomas-ham}. However, this strategy 
does not help in the case of the results presented here, and so is not
adopted in this paper.}.  This is the strategy we will take up
when we study the lorentzian theory.

\subsection*{Self-dual spacetimes and the deSitter solutions}

The equations of motion gotten from the Palatini action are,
\f
{\cal D} \wedge B =0  \ ,   \ \ \ \ \ \ \ \ \ \ \      F^i= \phi^i_j B^j
\ff

We can see immediately the self-dual solutions.    

\f
\Phi^i_j =- {\Lambda \over 3} \delta^i_j \ \ \rightarrow   
\ \ F^i= -{ \Lambda \over 3}  B^i
\ff

This shows us how the deSitter and self-dual solutions we
obtained from the Hamiltonian picture may be obtained directly
as solutions to the Euler-Lagrange equations.

\subsection*{Derivation of the Hamiltonian formalism}

Finally, we can briefly sketch how the constraints of the
hamiltonian that we guessed in section 2 are derived from the
forms of the action we have just described. It is easiest
to work with the $CDJ$ form of the action (\ref{CDJ}).

We begin by finding the canonical momenta, which is
\f
E^{ai}= \epsilon^{abc}F_{bc}^j (\phi^{-1})^i_j
\ff
with the canonical momenta for $A_0^i$, $E^{0i}$ of course vanishing.

The action can then be written as
\f
I^{CDJ}= \imath \int dt \int_\Sigma E^{ai}\dot{A}_{ai}- A_0^i {\cal G}_i
\ff
where the $\imath$ is there only for the Lorentzian case.
Because of this the Poisson brackets for the Lorentzian case have
the form eq. (\ref{lccr}) for the Lorentzian case and
eq. (\ref{ccr}) for the Euclidean theory.

The Gauss's law constraint (\ref{gausslaw}) then holds to preserve the
vanishing of  $E^{0i}$.  However there are additional constraints,
which arise from the fact that $\phi_{ij}$ is itself subject
to constraints, (\ref{constraint}), 
being symmetric and having trace fixed to be the
cosmological constant.  These must be
imposed to recover the equations of motion, because without them the
relationship between $\dot{A}_{ai}$ and the momenta cannot be inverted.

It is easy to check that the constraints that arise from the 
antisymmetric part of $\phi_{ij}$ vanishing is the diffeo constraint,
(\ref{diffeoconstraint}), 
while the constraint that arises from its trace being fixed is
the Hamiltonian constraint (\ref{hamconstraint}).  Thus we arrive at the
hamiltonian formulation we developed by guess work in section 2. 

\section{Boundaries with $\Lambda >0$ and Chern-Simons Theory}

There are several reasons we will want to consider spacetimes with
boundaries. These include the important subjects of how we realize 
the Bekenstein bound, study the entropy of black hole and cosmological
horizons and express the holographic principle. Depending on the context these
boundaries will be null, as at horizons, or timelike as in the
boundary of $AdS$ spacetimes or even Euclidean, if we are working in 
that
context. These boundaries may be at infinity, or they may have finite
area.  

Before we can study the quantum theory with boundaries we have to understand
the role boundaries play in the classical theory. Generally when 
there
is a boundary we will not have a sensible variational principle 
unless the
theory is modified to take the boundary into account. Normally these 
modifications consist of two parts. We have to add a boundary term to 
the action, which just depends on the fields pulled back into the 
boundary. And we have to add boundary conditions. Both the boundary 
action and boundary conditions must be chosen so that the variations of
the actions by the fields are pure bulk terms, so that the equations of
motion are sensible.

Here we will consider a region of spacetime with topology 
${\cal M} = \Sigma \times R$ with a boundary 
$\partial \Sigma = {\cal B}$.
We will study only one particular class of boundary conditions,
which are called {\it self-dual boundary 
conditions\cite{linking}. }

The basic idea of these boundary conditions is to require that at
the boundary some 
components of the fields satisfy the self-dual relations
(\ref{selfdual1}) which define the deSitter (or with $\Lambda <0$
anti-deSitter) spacetime\cite{linking}. We cannot require all the components
satisfy the self-dual conditions, otherwise only self-dual
solutions will be allowed. But we can get interesting boundary
conditions by requiring only a subset of the components
satisfy the self-dual relations (\ref{selfdual1}) when
pulled back into the boundary.

We will consider cases in which the spatial components of the
self-dual relation, pulled back into the boundary are satisfied,
in at least one spatial slicing of the boundary. Thus, we impose,
\f
F^i_{ab}|_{\cal B} =  -{\Lambda \over 3}
\epsilon_{abc} E^{ci}|_{\cal B} 
\label{CSboundarycondition}
\ff
There may be other components of the boundary condition, imposed on 
the timelike components of the boundary fields, for details about this 
in the euclidean case see \cite{linking}, in the timelike case 
see \cite{yime-holo} in the null 
case see \cite{kirill1,isolated}.  

To complement the boundary condition we must add a boundary term
to the action. The natural one to add turns out to be the
Chern-Simons action of the connection $A_a^i$ pulled back
into the boundary\cite{linking}. The action then reads, 

\f
I^{GR}= \epsilon \int_{\cal M}  B^{i}\wedge F_{i} 
+\phi_{ij}   B^{i}\wedge B^{j} + {\epsilon k \over 4\pi} \int_{\partial {\cal 
M}}Y_{CS} (A)
\label{CSboundaryterm}
\ff
where, from now on, $\epsilon=\imath$ for the Lorentzian theory and
unity for the Euclidean theory.  

There is in both cases a relation between the coupling constant, $k$ of
the Chern-Simons theory and the cosmological constant. 
\f
k={6\pi \over  \lambda} , \ \ \ \ \ \ \ \ \ \ \lambda = \hbar G \Lambda
\label{level}
\ff
\blankline

This ends our study of the classical physics we need to know
to understand the quantum theory of gravity with $\Lambda >0$.
The key lesson of this survey is the connection to topological
field theory, which we have seen arises three ways in the
classical theory:

\begin{itemize}
    
   \item{}The action for $GR$ has the form of a constrained 
   topological field theory.
   
   \item{}There is a natural class of boundary terms which require 
   that the boundary term added to the action is the Chern-Simons 
   action of the left handed spin connection, pulled back to the 
   boundary.
   
   \item{}The deSitter and other self-dual solutions follow gradients 
   of the Chern-Simons invariant, which can then be taken as the 
   Hamilton-Jacobi function.
   
   \end{itemize}

\part{The quantum theory}

\section{The Kodama State}

We begin with a very brief review of how diffeomorphism invariant
theories are to be quantized in the Hamiltonian approach.   For more details
on the basic approach, see \cite{tedlee,loop1,abhay-books}. 
We  do not here describe
path integral methods in loop quantum gravity, but they are well
developed, see, for example, \cite{mike-foam}-\cite{foamreviews}.

\subsection{A brief review of quantization}

The approach taken here is Dirac quantization. This means that
the whole unconstrained configuration space is quantized. 
This defines a {\it kinematical state space} $H^{kinematical}$.
The
constraints are imposed as operator relations on the states,
as in 
\f
\hat{\cal C} |\Psi > =0
\ff
where $\hat{\cal C}$ stands for operators representing all the first
class
constraints of the theory. The solutions to the constraints define
subspaces of the Hilbert space.  A physical state 
must be a simultaneous solution to all 
the constraints. 

Often this is done in two steps. The kernel of the
gauge and spatial diffeomorphism constraints is called
the diffeo-invariant Hilbert space, and is labeled $H^{diffeo}$.
The simultaneous kernel of all the constraints is called
the physical Hilbert space, $H^{physical}$.

Generally, new inner products need to be introduced on these Hilbert
spaces, because solutions to the constraints are not normalizable in
the inner products on the kinematical Hilbert space.  

We will work in this and the next four chapters with the connection 
representation of quantum gravity\cite{tedlee}.  After this we will
switch to the loop (or spin network) representation. As in the case
of the position and momentum representations in ordinary quantum 
mechanics these are equivalent, but complementary, in that certain
calculations are easier to do in one representation than another.

Both representations are defined as representations of a certain
algebra of classical observables. 

From a naive point of view, we could take the canonical commutation
relations (\ref{ccr}) as the basis of the quantization.    
Thus, we heuristically define
the connection representation by the relations
\f
<A|\Psi>= \Psi (A)  \ \ \  E^{ai}= 
 -\hbar G {\delta \over \delta A_{ai}}
\ff
These satisfy the commutation relations,
\f
[ A_a^i (x) , E^b_j (y) ] = \hbar G \delta_a^b \delta^i_j \delta^3 (y,x)
\ff
Note that because there is an $\imath$ in the classical commutation
relation (\ref{ccr}) no $\imath$ appears here\footnote{Were we working
instead with the Euclidean theory there would be an $\imath$ here.}.
Unless explicitly mentioned, from now on we are working with
the Lorentzian theory.

However, to discuss carefully the regularization of the operators 
that define the theory, we need to define the quantization in terms of
a different set of observables, which are the Wilson loops 
\f
T[\gamma, A]= Tr P e^{\int_\gamma A}
\ff
in the fundamental, spin $1/2$ representation, 
and the
elements of area, 
\f
A[{\cal S}]=\int_{\cal S} \sqrt{h}
\ff
where $\cal S$ is a surface in $\Sigma$ and $h$ is the determinant of
the induced metric in the surface.
These have very beautiful Poisson bracket relations,
\f
\{ T[\gamma, A], A[{\cal S}] \} = \hbar G Int [\gamma , {\cal S}]
T[\gamma, A]
\label{loopalgebra}
\ff
where $Int [\gamma , {\cal S}]$ is the intersection number of 
the loop and the surface. 

For the definitions of the connection and loop representations in 
terms of this algebra, see \cite{loop1}.  Here we will work with naive 
operators and mostly neglect the details of regularization procedures, which
can be found in the references. However, it is very important to 
stress that everything said here does go through when all the details 
of the regularization procedures are included. 

We now need the expressions of the constraints in the connection 
representation. These are

\begin{itemize}

\item{}Gauss's law:  
\f
{\cal G}^i \Psi (A) = {\cal D}_a {\delta \over \delta A_{ai}}\Psi 
(A) =0
\ff

\item{}Diffeomorphism constraint  
\f
{\cal H}_a \Psi (A) = F_{ab}^i{\delta \over \delta A_{bi}} \Psi (A)=0
\ff

\item{}Hamiltonian constraint:   
\f
{\cal H}\Psi (A)  = \epsilon_{ijk} {\delta \over \delta A_{ai}}{\delta 
\over \delta A_{bj}}
(  F^k_{ab} +  {\lambda \over 3} \epsilon_{abc}{\delta \over \delta A_{ck}} ) \Psi (A)
=0
\ff
\end{itemize}

Note that with the ordering given here, the quantum algebra of 
constraints can be shown, after a suitable regularization procedure, 
to be consistent\cite{abhay,tedlee,BGP,abhay-books}. 
This means that the commutators give terms 
proportional to operators, which are of the form of (new 
operator)$\times$ operator constraints. Thus, there are a non-trivial
space of states in the simultaneous kernel of all the constraints. 
In fact, infinite dimensional spaces of simultaneous solutions to all
the regularized constraints have been found and 
studied\cite{loop1,thomas-ham}.

For details of the different regularization procedures that can be
applied to define these constraints, and the infinite dimensional
families of solutions that have been found, see 
\cite{loop1,BGP,thomas-ham}.

For the time being, we will be concerned with 
the case that $\Sigma$ is compact, and without boundary. A bit
later we will show how boundaries are included in the quantum theory. 

\subsection{The Kodama state}

We will be concerned first of all with a particular simultaneous 
solution to all the constraints, which is the Kodama state we
introduced in section 4.  This is the Kodama state, defined 
by\cite{kodama}
\f
\Psi_{K} (A) =  {\cal N} e^{{3 \over 2 \lambda} \int Y_{CS}}
\ff

To show that this is a solution to all the constraints, one makes
use of the identity,
\f
  {\delta \Psi_K (A) \over \delta A_{ck}} = {3 \over  
2 \lambda} \epsilon^{abc}F_{ab}^i\Psi_K (A) 
\ff
Thus, the Kodama state is in the kernel of the operator
\f
{\cal J}_{ab}^i =  
F^k_{ab} +  {\lambda \over 3} \epsilon_{abc}{\delta \over \delta A_{ck}}
\ff
and satisfies the Hamiltonian constraint because we have
chosen an ordering such that
\f
{\cal H}  = \epsilon_{ijk} {\delta \over \delta A_{ai}}{\delta 
\over \delta A_{bj}} \cdot {\cal J}_{ab}^k
\ff
${\cal J}_{ab}^i$ is of course an operator version of the self-dual condition.

The Kodama state solves the other constraints because it is manifestly
invariant under diffeomorphisms of $\Sigma$ and {\it small}
gauge transformations. (Note that only small gauge transformations
are generated by constraints.)

One might think that invariance under large gauge transformations
would be achieved because $k$ is an integer.  However, this would
be wrong, as there is no $\imath$ in the Chern-Simons state. 

Invariance under large $SU(2)$ (real) gauge transformations is instead
gotten by choosing $\cal N$ to be a topological invariant also 
sensitive to them. For details of this, see the paper by Soo 
\cite{chopin-new}.  There is a reason for choosing $k$ to be an
integer, we will see it in section 11  below. 

It will also be important to note that as $A_a$ is complex, so is its
Chern-Simons invariant. Hence the Kodama state is complex.

The reader may now make the following queries:

\begin{itemize}

    \item{}{\bf Does the Kodama state survive the details of a 
    regularization procedure, needed to define the rigorous action
    of the constraints?} 
    
    {\it Yes, for details see \cite{BGP}.}
    
    \item{}{\bf Is the Kodama state normalizable?} 
    
    {\it Certainly it is not normalizable in the naive inner product.
    But this is to be expected, on two grounds. First because 
    solutions to constraints are generally never normalizable in the
    inner product of the kinematical Hilbert space\cite{abhay-books}. 
    Second,
    because, as we will see below, there are components of the 
    connection that function as a time coordinate on the configuration
    space\cite{chopinlee}. The physical inner product cannot integrate over 
    time, otherwise all energy eigenstates would be non-normalizable.
    Hence a new physical inner product needs to be chosen, and, given
    the fact that it satisfies all the physical properties we require
    of a physical state, it makes sense to take as a condition for
    the physical inner product that the Chern-Simons state, as well
    as its perturbations that we see below represent long wave graviton
    states, are normalizable.}

\item{}{\bf But, is the Chern-Simons state 
really a ground state? Does it really correspond to
the vacuum?}

{\it It is if one can study its weakly coupled excitations and they 
reproduce
quantum field theory in curved spacetime and long wave length
gravitons on deSitter spacetime. }

Note that we need only recover these for low energies in Planck units.

We show this first for matter, then for gravitons.

\end{itemize}

\section{The recovery of QFT on deSitter spacetime.} 

The first thing we can do to probe the Kodama state is to add
matter, and then see what happens if we excite the matter in the
presence of the state\footnote{The material in 
this section comes from ref. \cite{chopinlee} to which the
reader is referred for more details.}.

Adding matter fields is straightforward. In the language of loop quantum
gravity it is simple to add all kinds of matter: gauge fields, 
fermions, scalars, antisymmetric tensor gauge fields\footnote{There is also
no obstacle to extending the theory to supergravity, so long
as $\Lambda \leq 0$ in four dimensions\cite{super}.  This has
been worked out in some detail up to $N=2$ in four 
dimensions. For some partial results on $d=11$ supergravity,
the interested reader can see \cite{11d}.}.  
For what we are doing here
we do not need any details, so we will refer to all matter
fields as $\phi$, their canonical momenta as $\pi$
and the matter hamiltonian as $H^{matter}(\phi,\pi)$.

All the constraints get new terms in the matter fields. 
For the Hamiltonian constraint we have

\f 
{\cal H}^{total} = {\cal H}^{grav}(A,E) + 
H^{matter}(A,E,\phi,\pi)
\ff

We will work in an extended connection representation in which the
states are functionals $\Psi [A, \phi ]$, $\pi$ is represented
by $-\imath \hbar \delta / \delta \phi$ and so forth.

As in the pure gravity case, the gauge and diffeomorphism constraints,
applied to the states, require that the states are gauge invariant
and invariant under diffeomorphisms of $\Sigma$. 

This is straightforward, so we focus here on the hamiltonian
constraint. 

To study perturbations of the Kodama state,we follow the proposal of 
Banks\cite{banks1}, which is to study the semiclassical approximation in
quantum cosmology by a version of the Born-Oppenheim approximation, in 
which the gravitational degrees of freedom play the role of the heavy,
nuclear degrees of freedom, while the matter degrees of freedom play 
the role of the light, electron degrees of freedom. 

Thus, we consider a product state of the form
\f
\Psi (A,\phi) = \Psi_{K} (A) \chi (A, \phi )
\ff

The exact Hamiltonian constraint is then of the form
\f
\left (  H^{grav} + H^{matter} \right ) \Psi_K (A) \chi (A, \phi ) =0
\label{exactbanks}
\ff

The idea is to make an approximation to the exact equations,
which is described in terms of quantum matter fields propagating
on a classical background spacetime $(A^0, E^0)$. This approximation
is gotten by expanding the Wheeler-DeWitt equation (\ref{exactbanks})
in a neighborhood of a classical solution.  We use the fact that
the Kodama state can be understood as a $WKB$ state as well as
an exact solution. 
This tells us that the classical background
$(A^0, E^0)$ must be deSitter spacetime, as it is the unique
solution gotten by taking $S_{CS}$ to be the Hamiltonian-Jacobi
function, consistent with the requirement that the lorentzian metric
be real. 

We will describe the details of this approximation, for the case of a 
scalar field, in section 10. As a prelude, we mention here
the basic features of the results. 

As shown in \cite{chopinlee}, we find that an approximation to
(\ref{exactbanks}) takes the form of a Tomonaga-Schwinger equation:
\f
\imath {\delta \chi \over \delta \tau_{CS}} = 
 {1\over \Lambda} H^{matter}_{E^{ai}= (3/ \Lambda )  \epsilon^{abc}
F^i_{bc}} \chi  + O(l_{Pl}E)
\label{approx}
\ff
In this equation, the matter Hamiltonian is evaluated with classical 
    gravitational fields satisfying the self-dual condition
    $E^{ai}= (3/ \Lambda )  \epsilon^{abc} F^i_{bc} $.  As we just
    said, the reality conditions then tell us that the background is
    deSitter.   We have neglected higher order terms in
    $l_{Pl}E$, where $E$ is the energy of the matter fields measured
    with respect to the background metric. 
    
    The approximation procedure picks out a time coordinate called 
    $\tau_{CS}$, related to the Chern-Simons invariant. It is first 
    of all a 
    coordinate on the configuration space of the theory, defined by
    \f
    \delta \tau_{CS} (x) = { 1\over 2} {\cal I}m \ \epsilon^{abc}F_{bc}^{i}
    \delta A_{ai} (x)
    \ff
    Thus, integrated over the spatial manifold $\Sigma$, we have
    \f
    \int_{\Sigma} \delta \tau_{CS} (x) =  \delta {\cal I}m 
    \int Y_{CS}(A) 
    \ff
    If we take the integral, we can define
    \f
    T_{CS}= \int_\Sigma \tau_{CS}= Im \int_\Sigma Y_{CS}
    \label{CST}
    \ff
    This can be argued to provide a 
    provides a good global time parameter on the configuration 
    space\cite{chopinlee,chopinthesis}. This is because its derivative is always
    orthogonal, in the tangent space of the configuration space, to 
    both the gauge directions and the directions that parameterize the
    physical degrees of freedom.  
    
    When evaluated on a background solution, this gives rise to a time
    coordinate on the spacetime. One can then show that, to leading
    order in $\lambda$, $Y_{CS}= \imath \sqrt{det(q)}{\cal K} + O(\sqrt{\lambda})$,
   where ${\cal K}$ is the trace of the extrinsic curvature $K_{ab}$.
     Thus. this choice of time coordinate agrees, to leading order in 
     $\lambda$, with that proposed by York\cite{chopinlee}.  
      This 
    time coordinate has been shown to have many good properties that 
    an intrinsic time coordinate should have.

Thus, QFT on deSitter {\it is} a good approximation to the 
physics of   $\Psi (A,\phi) = \Psi_K (A) \chi (A, \phi )$ 
when $ \lambda=\hbar G \Lambda$ and $l_{Planck}E$ are small. 
This stands as a first piece of evidence that $\Psi_K (A)$ may
be indeed a good ground state.  

There are additional terms in $l_{Planck}E$, where $E$ is the 
   matter energy on 
      the deSitter background.
      We will study the effect of these terms in section 10.

\section{Gravitons from perturbations around the Kodama state}

To further probe the properties of the Kodama state we should also 
investigate its gravitational excitations. To do this we return to the 
case of pure gravity and consider states of the form
\f
\Psi [A]={\cal N} e^{{3\over 2\lambda} \int Y_{CS} +\lambda  S^{\prime} (A)}
\label{gproduct}
\ff

It is not difficult to show that there are solutions of this form, and that they
do describe long wavelength gravitons moving on the classical 
background of deSitter spacetime\footnote{More details concerning
these results will be reported elsewhere\cite{inprogress,newcarlolee}.}. But 
to do this we first need to know
how to recognize gravitons in this language. We then detour to 
summarize the results in this area,

\subsection*{Linearized gravity on a deSitter background}

The quantization of linearized general relativity in Sen-Ashtekar 
variables of the kind we are using was considered early in the study
of loop quantum gravity\cite{gravitons}. 
There a complete description 
was obtained of gravitons on a Minkowski spacetime background. 
It is trivial to extend
what was done there to gravitons moving on a deSitter background. 
As we want results that hold for small $\lambda$, it is convenient
to use $\lambda$ as an expansion parameter. 

Thus, we expand classical general relativity around the deSitter
background studied in section 3
\f
A_{ai}= \imath \sqrt{\Lambda}f(t) \delta_{ai} + \lambda a_{ai}; \ \ \
E^{ai}= f^{2}\delta^{ai} + \lambda e^{ai}
\ff

It is trivial then to compute the constraints to linear order
to find the
linearized constraints satisfied by the $a_{ai}$'s and
$e^{ai}$'s.
To solve them we need to impose $7$ gauge fixing conditions.
A natural set to impose 
is\footnote{As an exercise one has to check that these seven gauge fixing 
conditions do together with the linearized constraints make a second
class algebra.}
\f
a_{[ai]}= a_{a}^{a}= \partial_{a}a^{a}_{i}=0
\ff
where indices are raised and lowered by the background metric, 
$q_{ab}^{0}$.
The simultaneous solution of the linearized constraints and gauge
fixing conditions is
\f
\partial_{a}e^{a}_{i}=e_{[ai]}= e_{a}^{a} = 0 .
\ff
The result is that the theory is reduced to $a^{r}$'s and $e^{r}$'s
that are tracefree, divergence free and symmetric. These are
spin two fields. 

The linearized poisson brackets can be derived by applying the
full Poisson brackets to the linearized fields. This gives,
\f
\{ a_{ai}^{r}(x) , e^{bj}_{r}(y) \} = \imath 
P_{ai}^{bj} \delta^{3} (x,y)
\ff
where $P_{ai}^{bj}$ is the projection operator onto the symmetric,
transverse, tracefree fields. 

Finally, we have to construct the linearized hamiltonian. This comes
from the quadratic terms in the integral of the hamiltonian constraint,
and comes out to be, 
\f
h (a^{r},e^{r}) = f^{-1}\left [ 
\epsilon^{ajk}({\cal D}^{0}_{a} a_{bk}^{r})e^{b}_{r \ j}
+\Lambda e^{ai}_{r}e_{r \ ia}
\right ]
\ff

It is then straightforward to quantize this theory, yielding a quantum 
theory of gravitons on the background of deSitter spacetime.

\subsection*{Linearization of the exact quantum theory agrees with the 
quantization of the linearized theory, for long wavelength.}

Now we want to go the other way around and study expansions of exact 
states in powers of $\lambda$ around the Kodama state.  We consider a 
product state of the form (\ref{gproduct}) and  solve
all seven constraints in a neighborhood of the classical trajectory
on the configuration space.  

The 6 kinematical constraints  give:
\f
\int_{\Sigma} ({\cal D}_{a} w)^{i}{\delta S^{\prime}\over \delta A_{ai}} =0 ; 
\ \ \
\int_{\Sigma} v^{a}F_{ab}^{i}{\delta S^{\prime}\over \delta A_{ai}} =0
\ff
where $w^{i}$ and $v^{a}$ are arbitrary functions on $\Sigma$. These
are linearized around the $dS$ background.  
They are solved by taking $S^{\prime}=S^{\prime}(f, a_{ai})$ with  $a_{ai}$
symmetric and transverse wrt the dS background. 
Thus, 
\f
{\delta S^{\prime}\over \delta A_{ai}}=
{\imath \over \sqrt{\Lambda}} \delta_{ai}
{\delta  S^{\prime} \over \delta f}
+{1\over \lambda} {\delta  S^{\prime} \over \delta a_{ai}}
\label{deltaA}
\ff

The hamiltonian constraint is,
\f
\epsilon_{abc}\epsilon_{ijk} {\delta \over \delta A_{ai}} 
{\delta \over \delta A_{bj}}  \left [
{\delta S^{\prime}\over \delta A_{ck}}  
e^{{3\over 2\Lambda} \int Y_{CS} + S^{\prime} (A)}
\right ]=0
\ff

Using (\ref{deltaA}),
this can be expanded to give:
\f
\imath {\partial S^{\prime}\over \partial t} = 
\hat{H}^{2}   S^{\prime}+ O(l_{Planck}E) + O(\sqrt{\lambda} )
\ff

where the free Hamiltonian is

\f
\hat{H}^{2}= \hat{h}(\hat{a},\hat{e}={\delta \over \delta a})
\ff
and $E$ is the energy of the graviton state with respect to the
background.
Thus we conclude that 
for long wavelength perturbations, but only so long as
    $l_{Pl} E << 1$, 
    the linearized theory is recovered.  However it must be 
    emphasized that we have only obtained a correspondence with the 
    standard linearized theory for low energy and small
    $\lambda$.   
    
    To go beyond this, which will be necessary if we want to compute
    graviton graviton scattering, we need to have exact expressions
    for the physical states which are obtained by local perturbations
    of the Kodama state.  This requires solving the kinematical 
    constraints exactly, rather than order by order in $l_{Pl}$
    and $\lambda$. To do this we must go to the loop representation.
    This will be the subject of sections 12 and 13. However,
    before going there there are a few more things of physical
    interest we can learn from the connection representation.

\section{Corrections to energy momentum relations}

Now that we have recovered known physics from the Kodama
solution, we may go on to see if the 
theory makes any predictions beyond the recovery of quantum field
theory in the semiclassical limit.  To see that it may, let us
look in detail at the fundamental equation (\ref{exactbanks}).
For simplicity we consider the case of a massless, non-interacting
scalar field, although
similar conclusions apply for other matter fields\footnote{More details
concerning the results of this section will appear in \cite{newcarlolee}.}.  
For this case the
form of the matter term in the Hamiltonian constraint is 
\f
H^{matter} (x) = {G \hbar \over 2} \left ( \pi^{2} + 
 (\partial_{a}\phi )(\partial_{b}\phi )
E^{ai}E^{b}_{i}
\right )
\ff
where $\pi$ is the canonical momentum of the scalar field. 
Implemented as a quantum operator this is,
\f
\hat{H}^{matter}(x)\Psi_{K}(A) \chi (A,\phi) = {\hbar G\over 2}
\left ( \pi^{2} + (\hbar G)^{2} (\partial_{a}\phi )(\partial_{b}\phi )
{\delta \over \delta A_{ai}}{\delta \over \delta A_{b}^{i}}
\right ) \Psi_{K}(A) \chi (A,\phi)
\label{matter}
\ff
In section 8 we recovered quantum field theory in curved spacetime
from an approximation to this last expression. In this approximation
we considered only the terms in which 
the factors of  
$\hat{E}^{ai}= -\hbar G \delta /\delta A_{ai}$ in the
second term act on the Kodama state, giving terms proportional
to the background frame field , $\delta_{ai} f^{2}$. 
Keeping only these terms we have 
\f
\hat{H}^{matter}(x)\Psi_{K}(A) \chi (A,\phi) = {1\over 2}
\left ( \pi^{2} + \delta^{ab} f^{4} (\partial_{a}\phi )(\partial_{b}\phi )
\right )  \chi (A,\phi)
\ff
which is the hamiltonian for the scalar field on the background
spacetime. 

To go beyond the semiclassical approximation we may then
consider the other terms in (\ref{matter}) in which one or both of the
functional derivatives by $A_{ai}$ act directly on the perturbed state
$\chi (A,\phi)$. These still give terms linear in $\chi$
so they may be interpreted as corrections to
the functional Schroedinger equation.  
We will see that these terms give predictions of new physics.

The interpretation of the new terms is easiest when we can neglect
the effect of the cosmological constant, and approximate a region
of deSitter spacetime by a region of flat spacetime. 
To get predictions for the theory in flat spacetime, we proceed
in two steps. First we
evaluate the approximate solutions to the Wheeler
deWitt equation at the background values of the connection
and metric we studied in section 3. These are given by eqs. 
(\ref{Abackground},\ref{Ebackground}).  We then approach
flat spacetime by neglecting terms such as $k^{2}\Lambda$, 
where $k$ is the momentum of a particle, which vanish in the
limit $\Lambda \rightarrow 0$.  This is of course a good
approximation in the observed situation in which 
the cosmological constant is non-zero, but very small.

Evaluating
the action of $\delta /\delta A_{ai}$ to leading
order on $\Psi_{K}(A)$, we found that, 
\f
\hat{E}^{ai}(x)\Psi_K (A) =  
-\hbar G {\delta \Psi_K (A) \over \delta A_{ai}(x)} =
 f^{2}(t) \delta_{ai} \Psi_K (A)
 \label{Kaction}
\ff
Thus, the full action of the functional derivatives gives
\begin{eqnarray}
{(\hbar G)^{2 }\over 2} (\partial_{a}\phi )(\partial_{b}\phi ) 
\left (
{\delta \over \delta A_{ai}}{\delta \over \delta A_{b}^{i}}
\right ) 
\Psi_{K}(A) \chi (A,\phi) &= & \Psi_{K}(A) \left \{ 
{ f^{4}\over 2} 
(\partial_{a}\phi )(\partial_{b}\phi ) \delta^{ab} \chi (A,\phi) \right. 
 \\
&+& \left.  (\partial_{a}\phi )(\partial_{b}\phi ) [ -\hbar G f^{2}
{\delta \chi (A,\phi ) \over \delta A_{ab}}  + {(\hbar G)^{2} \over 2}
{\delta^{2} \chi (A,\phi ) \over \delta A_{ai} \delta A_{b}^{i}}  ]
\right \}  \nonumber
\end{eqnarray}

In section 8 we also saw that the 
time derivative in the functional Schroedinger equation
came from terms in which a single derivative in ${\delta \over \delta A_{ai}}$
in the gravity part of the hamiltonian
constraint acted on $\chi (A,\phi )$, while the remaining
functional derivatives act on $\Psi_{K} (A)$ giving 
factors of the background fields through (\ref{Kaction}). 
But there are also terms
in the gravity part 
in which two and three functional derivatives act on $\chi (A,\phi )$.
This will give additional corrections to the functional Schroedinger
equation.

Before writing them all out we have to consider the
effect of the gauge and diffeomorphism constraints. 
By an analysis similar to the one of the last section, they
tell us that
\f
\chi(A, \phi ) = \chi (\tau , a_{ai}, \phi )
\ff
where as before $a_{ai}$ is transverse and tracefree. 
The dependence on the $a_{ai}$ describes the couplings
to gravitons.  

$\tau$ is a field that parameterizes the trace part of the background
$A_{ai}$ and is given by
\f
A_{ai}(x) = \delta_{ai}e^{\sqrt{\Lambda \over 3} \tau (x)} + \ldots
\ff
For the background $A_{ai}$ discussed in section 3 we have
$\tau (x) =t$. However its important to keep the distinction
clear: while $t$ is a coordinate on a particular classical
solution, $\tau (x)$ is a field that parameterizes the
trace part of $A_{ai}$ on the whole configuration space. 

On the solution, $\tau$ is related to the Chern-Simon time
described in section 8 by
\f
\tau_{CS}(x) =\left ({\Lambda \over 3}  \right )^{3/2}
e^{3\sqrt{\Lambda \over 3}\tau(x)}
\ff

Thus, we have 
\f
\hat{E}^{ai}\chi (A,\phi )=
-\hbar G {\delta \chi (A,\phi ) \over \delta A_{ai}}=
{\imath \hbar G \over {\Lambda}} \delta_{ai}
{\delta  \chi (A,\phi )  \over \delta \tau}
- \hbar G {\delta  \chi (A,\phi ) \over \delta a_{ai}}
\ff
We are interested in finding leading order corrections
to the propagation of a free field on the background
spacetime.  Thus, we can neglect
the couplings to gravitons.  Doing so gives us corrections
to the 
Tomonaga-Schwinger equation: 
\begin{eqnarray}
\imath f^{4}{\delta \chi (\tau, \phi ) \over \delta \tau (x) } 
&=& {1\over 2} [ \pi^{2}+f^{4}(\partial_{a}\phi )^{2}]  \chi (\tau, \phi )
\nonumber \\
&&+ {1\over 2} (\partial_{a}\phi )^{2} \left [
{2 \imath \hbar G f^{2} \over \Lambda} {\delta  \over \delta \tau (x) }
- \left ( { \hbar G \over \Lambda} \right )^{2}
{\delta^{2}  \over \delta \tau^{2} (x) }
\right ]  \chi (\tau, \phi )  
\nonumber \\
&&+ \left [ 2 { \hbar G f^{2} \over \Lambda}  
{\delta^{2}  \over \delta \tau^{2} (x) }
+ \imath \left ( { \hbar G \over \Lambda} \right )^{2}
{\delta^{3}  \over \delta \tau^{3} (x) }
\right ] \chi (\tau, \phi ) 
\end{eqnarray}

The corrections on the second line come from the action of
$\delta \over \delta A_{ai}$ on $\chi$ from the matter hamiltonian
density, while the corrections on the last line come from
the higher order terms in $\delta \over \delta A_{ai}$ from
the gravitational part of the hamiltonian constraint.  

To see what the effect of the corrections is on ordinary physics, we
have to re-express the Tomonoaga-Schwinger equation in terms of
measurable quantities that govern the low energy physics.
One way to approach this is the following. 

We are interested in extracting quantum field theory on Minkowski
spacetime, in the limit $\Lambda \rightarrow 0$.  For the limit
to be non-singular we must rescale the time coordinate, 
because of the factors of $\hbar G / \Lambda$ in front of
the $\delta /\delta \tau$ derivatives. In any case we 
need to rescale to remove a density factor, as we are interested
in expressing the final answer in terms of a Schrodinger equation
rather than a Tomonaga-Schwinger type equation. To do this we
must replace the functional degree of freedom $\tau (x)$, which 
we have chosen to represent time by a global coordinate $T$.
This coordinate $T$ is taken to be proportional to $\tau$ on
a $\tau=$ constant slice. However $\delta /\delta \tau (x)$
and $\partial /\partial T$ have different density weights
and dimensions and this must be compensated for.

We accomplish both if 
we rescale so that on a fixed $\tau =$ constant slice, 
\f
{ \hbar G \over \Lambda} {\delta  \over \delta \tau (x) } = 
\alpha l_{Pl} \sqrt{detq_{ab}^{0}}{\partial \over \partial T}
\ff
where $\alpha$ is a dimensionless parameter.  The factor of
$\sqrt{detq_{ab}^{0}}$ is due to the fact that 
$\delta /\delta \tau (x)$ is a density.
This form is required
as $l_{Pl}$ is the only dimensional parameter in the theory
when $\Lambda \rightarrow 0$.  We will see shortly how $\alpha$ is 
fixed.  

The next step is to integrate over the spatial manifold, so
as to recover the Schroedinger equation.  To do this we multiply
the whole expression by $1/\sqrt{detq_{ab}^{0}}$, because the
form of the hamiltonian constraint we are using has density
weight two, and then integrate. We set $f=1$ as we are about to
take $\Lambda \rightarrow 0$ and we note that in the coordinates
we are using $det(q_{qb}^{0})=1$.
This gives us,
\begin{eqnarray}
\imath {\partial \chi (T, \phi ) \over \partial T  } 
\left ( {\alpha V \Lambda \over l_{Pl}}  \right )
&=& \int_{\Sigma} {1\over 2} [ \pi^{2}+(\partial_{a}\phi )^{2}] 
\chi (T, \phi )
\nonumber \\
&&+ \int_{\Sigma} {1\over 2} (\partial_{a}\phi )^{2} \left [
2 \imath \alpha l_{Pl} {\partial \over \partial T  } 
-  \alpha^{2} l_{Pl}^{2}
{\partial^{2}¥ \over \partial T^{2}  }
\right ]  \chi (T, \phi ) 
\nonumber \\
&&+\left ( {\alpha V \Lambda \over l_{Pl}}  \right )
\left [
2\alpha^{2}l_{Pl}^{2}{\partial  \over \partial T  }
+ \imath \alpha^{3} l_{Pl}^{3}{\partial^{2}  \over \partial T^{2}  }
\right ] \chi (T, \phi )
\end{eqnarray}
where $V$ is the volume of the spatial manifold  according to the
background metric,
$V= \int_{\Sigma}\sqrt{det(q^{0})}$.  We impose an infrared cutoff
so $V$ is finite. We will shortly take $V\rightarrow \infty $ as
$\Lambda \rightarrow 0$. 

However before we do this we should take into account the 
renormalization between the bare fields and the physical fields
that enter into the low energy physics. 
We expect
to have to renormalize because there are interactions between the
scalar and gravitational fields. However, as there is a cutoff on
the spatial resolution in the exact diffeomorphism invariant
states\footnote{See the sections on the loop representation.}
we expect the wavefunction renormalization to be finite and to be
proportional to powers of the ratio $L \over l_{Pl}$, with
$V=L^{3}$ as they represent infrared and ultraviolet cutoffs.
Further as we expect relativistic invariance to hold at least
up to corrections in $l_{Pl}$, we expect that 
$\pi \approx \dot{\phi}$ and $\partial_{a}\phi $ to renormalize
by the same factor, again up to possible corrections in
$l_{Pl}$\footnote{In the following we ignore such corrections,
but if found by calculations they can be inserted directly in the
following expressions.}. 
Thus we expect
\f
\pi= Z \pi_{R} , \ \ \ \ \ \  \partial_{a}\phi = Z \partial_{a}\phi_{R}
\ff
where $Z$ is a multiplicative renormalization.
Let us suppose that $Z= \beta^{{1/2}} (  L/l_{Pl})^{d/2}$.  As the background
represents deSitter spacetime, it is natural to scale $\Lambda = \gamma /L^{2}$
where $\gamma$ is a factor of order one depending on the topology.
Thus we have
\begin{eqnarray}
\imath {\partial \chi (T, \phi ) \over \partial T  } 
{\alpha  \gamma \over \beta} \left (  { l_{Pl} \over L} \right )^{(d-1)}
&=& \int_{\Sigma} {1\over 2} [ \pi^{2}_{R}+(\partial_{a}\phi_{R} )^{2}] 
\chi (T, \phi )
\nonumber \\
&&+ \int_{\Sigma} {1\over 2} (\partial_{a}\phi_{R} )^{2} \left [
2\imath \alpha l_{Pl} {\partial \over \partial T  } 
-  \alpha^{2} l_{Pl}^{2}
{\partial^{2} \over \partial T^{2}  }
\right ]  \chi (T, \phi ) 
\nonumber \\
&&+ {\alpha  \gamma \over \beta} \left (  { l_{Pl} \over L} \right )^{(d-1)}
\left [
2\alpha^{2}l_{Pl}^{2}{\partial^{2}  \over \partial T^{2}  }
+ \imath \alpha^{3} l_{Pl}^{3} {\partial^{3}  \over \partial T^{3}  }
\right ] \chi (T, \phi )
\end{eqnarray}
We must recover the Schroedinger equation in the limit
$L \rightarrow \infty $, $l_{Pl}\rightarrow 0$. As the 
renormalized Hamiltonian
\f
H_{R}= \int_{\Sigma} {1\over 2} [ \pi^{2}_{R}+(\partial_{a}\phi_{R} )^{2}] 
\ff
should generate evolution in $T$, we require that the coefficient of
$\imath {\partial \over \partial T}$ on the left hand side be unity.
This tells us that
\f
\alpha ={ \beta \over \gamma } \left ( {L \over l_{Pl}} \right )^{(d-1)}
\ff

The limit then exists for $d \leq 1$. 
If $d<1$ the additional terms disappear, and the usual Lorentz
invariant quantum field theory is recovered. But in
the case that $d=1$   
we have,
\f
\alpha = {\beta \over \gamma }
\ff
is a factor of order unity. Then our equation is
\begin{eqnarray}
\imath {\partial \chi (T, \phi ) \over \partial T  } 
&=&  H_{R}
\chi (T, \phi )
\nonumber \\
&&+ \int_{\Sigma} {1\over 2} (\partial_{a}\phi_{R} )^{2} \left [
2\imath \alpha l_{Pl} {\partial \over \partial T  } 
-  \alpha^{2} l_{Pl}^{2}
{\partial^{2} \over \partial T^{2}  }
\right ]  \chi (T, \phi ) 
\nonumber \\
&&+\left [
2\alpha^{2}l_{Pl}^{2}{\partial^{2}  \over \partial T^{2}  }
+ \imath \alpha^{3} l_{Pl}^{3}{\partial^{3}  \over \partial T^{3}  }
\right ] \chi (T, \phi )
\label{correctedS}
\end{eqnarray}
Thus, under the assumptions stated, we predict corrections to the Schroedinger equation,
of order $l_{Pl}$, with the finite dimensionless coefficient $\alpha$
determined by the wavefunction renormalization of the scalar field
theory interacting with gravity. 

Now, to analyze the scalar field theory 
we can use to a first approximation a regular Fock space quantization in which
\f
\phi (x,t) = \int {d^{3}k \over \sqrt{(2\pi)^{3}2\omega}} \left [ 
a_{k} f_{k}(x)e^{-\imath \omega t} + 
 a_{k}^{\dagger} f_{k}^{*}(x)e^{\imath \omega t}
\right ]
\label{field}
\ff
where, as $\Lambda \rightarrow 0$, $f_{k}(x) \approx e^{\imath 
k_{a}x^{a}}$.

The Fock space one particle states are not going to be exact
solutions to the Wheeler-DeWitt equation, but we can search
for solutions of the form
\f
\chi (T, \phi ) = e^{-\imath \omega T} |k> + O(l_{Pl})
\ff
where $|k>$ is a one particle Fock state.
We do not set $\omega = |k|$ in (\ref{field}), instead we
take the components $<k|\ldots|k> $ of the 
Wheeler-DeWitt equation to extract the relation between
$\omega$ and $|k|$. 

We find, after the standard normal ordering
\f
<k|:H_{R}: |k> = {\omega^{2}+ k^{2} \over 2\omega}
\ff
\f
<k| :\int_{\Sigma }(\partial_{a}\phi_{R})^{2}:|k> ={k^{2}\over 2\omega}
\ff
Applying eq. (\ref{correctedS}) to this state we find
\f
\omega^{2}{ \left ( 1+ 4\alpha l_{Pl}\omega + 
2\alpha^{2}l_{Pl}^{2}\omega^{2} \right ) \over 
\left ( 1+ \alpha l_{Pl}\omega + {1\over 2}
\alpha^{2}l_{Pl}^{2}\omega^{2} \right )}
= k^{2}
\ff

We thus see that there are corrections to the energy momentum
relations. These kinds of corrections have been
discussed by a number of authors, and with $\alpha$ of order unity
are expected to be measurable in experiments involving gamma
ray busts and cosmic rays. (For details see 
\cite{GAC1,AC-Piron,seth,testreviews}.)  
For example we predict that a massless scalar particle travels
with a speed
\f
v= {d\omega \over d|k|} = 1-{3\over 2} \alpha l_{Pl}\omega +\ldots
\ff
If the same effect occurs for photons it is expected in the near future
that it will be detectable
in timing experiments in gamma ray bursts\cite{GAC1,testreviews}.  
It is straightforward to extend these results to photons
and other fields, these results will be reported elsewhere. 

It is very interesting to note that similar corrections have been derived
by Gambini and Pullin\cite{GP} and by 
Alfaro, Morales-Tecotl and Urrutia\cite{AMU} 
from loop quantum gravity, using
a different approach in which they study the propagation of
matter fields on background spatial geometries represented
by ``weave states.''  They concluded that the exact value of the coefficient
$\alpha$ depends on the details of the wavefunction of the ground 
state, here, having chosen a particular groundstate we 
find a prediction for $\alpha$ involving also the wavefunction
renormalization of the matter field. 

These corrections raise several intriguing issues, which are
studied in some detail in \cite{newcarlolee}.   Among the
issues they raise is that of Lorentz invariance.  Of course one
possibility is that Lorentz invariance is simply broken.  A possible
source of the breaking of Lorentz invariance in this case is that we
studied the Wheeler DeWitt equation in the neighborhood of the
connection described in section 3. 
This gives the spatially flat coordinatization
of a region of deSitter spacetime, where the region coordinatized is 
the interior of a horizon of a single inertial observer. 

However, we might have considered a semiclassical approximation to the 
Wheeler-deWitt equation corresponding to a global coordinatization
of deSitter spacetime. Alternatively, we might have considered 
expanding around a semiclassical approximation to a coordinatization
of the region accessible to  any inertial 
observer in deSitter spacetime. Thus, we might
have expected relativistic invariance to be restored in the limit
$\Lambda \rightarrow 0$.

To study this issue we must go beyond the approximation of
quantum field theory on fixed backgrounds, as the quantum
fields defined in different causal regions of a fixed spacetime
background are generally not unitarily equivalent. This raises a host
of issues, but they should be resolved by the full quantum theory
of gravity. As we have in fact been studying a well defined 
approximation to the full theory, this should in the future be possible.

We may note that one possible outcome of such a study is that 
the principle of the relativity of inertial observers is maintained,
but the lorentz transformations become non-linear when energies
and momenta are of the order of the Planck scale. This possibility
has been studied by a number of 
authors\cite{gac-dsr,dsr2,joaolee1,joaolee2}. We may note that
the deformed energy momentum relation we arrived at here is of
a form that would be permitted in such a modified formulation
of the Lorentz transformations\cite{joaolee2}.

\section{The thermal nature of quantum gravity 
with $\Lambda > 0$.}

It is well established that quantum field theory on
deSitter spacetime must be interpreted as irreducibly thermal\cite{GH},
with a temperature given by 
\f
{\cal T} = { 1\over 2\pi} \sqrt{\Lambda \over 3}
\label{dShot}
\ff
This can be understood as due to the presence of the horizon.
Alternatively, one can show that any quantum field on deSitter
spacetime, satisfies the $KMS$ condition for a thermal state.
This is that the continuation of any correlation function to
imaginary time coordinate $t_{E}= \imath t$ be periodic,
with period $\beta= 1/{\cal T}$.  

What is not so well known, however, is that in loop quantum
gravity the full background independent quantum theory of gravity
plus arbitrary matter fields has also an irreducibly thermal nature.
This is because it satisfies the $KMS$ condition on the whole
configuration space\footnote{The argument of this section is 
taken from \cite{chopinlee}.}.

To apply the $KMS$ condition to quantum gravity we need two things:
1) a definition of a time coordinate on the configuration space
and 2) a definition of the
continuation to Euclidean time. In a background independent theory 
we cannot use a time
coordinate on a given classical spacetime, as that is just a given
classical solution. We need instead a time coordinate on the
configuration space of the theory.   

We saw above in section 8 that there is in fact
a preferred time coordinate  on the configuration space,
which is picked out by the semiclassical expansion around
the Kodama state. It is equal to the imaginary part of the
Chern-Simons invariant, eq. (\ref{CST}).  There are other arguments
that confirm that (\ref{CST}) is a good time coordinate on the
configuration space, for example it is always normal to
the gauge directions in the tangent space of the configuration
space. For details see \cite{chopinlee,chopinthesis}. 
The Chern-Simons time coordinate is dimensionless, as we saw
in the last sections when we evaluate it on a given solution
we have to scale it appropriately.  

It is interesting to note that 
 (Lorentzian) Kodama state can be written
\f
\Psi_{K}(A)= e^{\imath M T_{CS}}e^{{k\over 4\pi}{\cal R}e \int Y_{CS}(A) }
\ff
where the dimensionless ``energy'' is
\f
M={ k \over 4\pi} = {3 \over 2 \lambda}
\ff

Now we need a definition of how to continue to Euclidean time.  As
we are dealing with a theory of spacetime we should continue the 
whole theory to Euclidean signature. This requires the following
changes: The connection $A_{ai}$ becomes a real, $SU(2)$ connection 
and 
there is now an $\imath$ in $E^{ai}=\imath \delta /\delta A_{ai}$.
As a consequence of which the Chern-Simons state is now,
\f
\Psi_{K}^{Euc}(A)= e^{{\imath k\over 4\pi} \int Y_{CS}(A) }
\label{CSEuclid}
\ff
The Euclidean time coordinate is then just
\f
T_{ECS}=  \int Y_{CS}(A)
\label{ECS}
\ff
as can be seen directly, or by repeating the derivation
from the semiclassical theory.  Thus, the Euclidean
wavefunction is,
\f
\Psi_{K}^{Euc}(A)= e^{\imath {k\over 4\pi} T_{Euc}  }
\ff
This  is periodic
in $T_{Euc}$.  
However, this is not enough to show that the $KMS$ condition
is satisfied, for that requires that {\it every correlation function}
be periodic. 

Interestingly enough, this is in fact the case whenever
$\Sigma$ is chosen so that $\pi^{3}(\Sigma)$ is nontrivial.
In this case there are large gauge transformations that have
the property that

\f
\int Y_{CS}(A) \rightarrow \int Y_{CS}(A) + 8 \pi^{2}n
\ff
where $n$ is the winding number of the large gauge transformation.
This means that $T_{ECS}=\int Y_{CS}(A)$ is actually
a periodic function on the configuration space. As a result,
{\it every} correlation function will satisfy the $KMS$ condition
in $T_{ECS}$, no matter what the state.  That is, by equating 
configurations of $A_{ai}$ that differ by a  
large gauge transformations we reduce the topology of the
configuration space to a circle, which is parameterized by $T_{ECS}$.  

As a result of this universal periodicity there is a temperature,
given in dimensionless units by 
${\cal T}_{dimless}= {1\over 8 \pi^{2}}$.  This dimensionless 
temperature corresponds to the fact that the time coordinate
on the configuration space, $T_{CS}$ is dimensionless.

It is intereseting to ask if this dimensionless temperature 
corresponds to the temperature on deSitter spacetime.  To
investigate this we may consider a trajectory in 
configuration space that corresponds to a slicing of deSitter
spacetime with topology $S^{3}\times R$. Such coordinates are
given by 
\f
ds^{2}= -(1 -{ \Lambda r^{2}\over 3}) dt^{2} +
{1 \over (1 -{ \Lambda r^{2}\over 3})} dr^{2} + d\Omega^{2}
\label{solution}
\ff
To work out the scaling of the coordinate $t$ on the
solution with the coordinate $T_{CS}$ on the configuration
space, we compute
\f
{\partial T_{CS} \over \partial t} = \int_{S^{3}} N 
\{ T_{CS}(A), {\cal H} \} 
\ff
where the (densitized) lapse $N$ is read off from the
solution (\ref{solution}).  A simple calculation gives
\f
{\partial T_{CS} \over \partial t} = 4\pi \sqrt{\Lambda \over 3}
\ff
Thus, if the Euclidean continuation $T_{ECS}$ is periodic
with period $8 \pi^{2}$, the Euclidean continuation of the
time coordinate on the solution must be periodic with
period $2\pi \sqrt{3 \over \Lambda}$. In fact, this is the
periodicity of the Euclidean deSitter solution, in these 
coordinates! This leads to the temperature of deSitter
spacetime, (\ref{dShot}).

Thus, we learn that {\it the periodicity of the
Euclidean deSitter spacetime is a consequence of that
spacetime having an interpretation as a 
trajectory on the configuration space of
$SU(2)$ connections.} The periodicity of the Euclidean
Schwarzschild solution is a consequence of the fact that the whole
configuration space is periodic
due to the action of the large gauge transformations.  This is yet
another connection between the properties of the gauge theory
and the physics of gravitation.  Thus, 
the thermal nature of quantum field theory on deSitter
spacetime is a consequence of a deeper and more general result, 
which is that 
the whole quantum theory with $\Lambda >0$ is thermal.

Finally, we can deduce one more fact from these considerations.
For the analysis we have just given to be relevant to the Kodama
state, it must be that the 
Euclidean Kodama state is itself well defined under large gauge
transformations. This will only be the case if $k$ is an integer.

\section{The loop transform for positive $\Lambda$}

To go beyond a semiclassical expansion we need to study the exact
quantum states of the theory. This can be done using the loop,
or as it is sometimes called, the spin network, representation of
the state space. Here we discuss the basics of the loop
representation, paying special attention to the modifications
required for $\Lambda \neq 0$.  

The loop representation can be defined in 
different ways, all leading to the same results.
The most secure way is to construct it directly as a representation
of the loop algebra, equation (\ref{loopalgebra}).  Alternative constructions
using rigorous methods involving measures on the space of connections
are described in \cite{thomas-thesis}.  Here we will use the
original technique, which is to construct the loop representation
from the connection representation by means of a linear transform
called the loop transform\cite{loop1,loop2}.

The basic form of the transform is

\f
\tilde{\Psi} (\Gamma ) = \int d\mu (A) T[\Gamma , A]  \Psi (A) 
\ff
where $\Gamma$ is a loop, or set of loops, $\gamma_i$, in which
case $T[\Gamma , A]= \prod_i T[\gamma_i , A]$, where $T[\gamma_i , A]$
is the trace of the holonomy of the connection around the
loop $\gamma_{i}$ in the spin $1/2$ representation.  Finally, $d\mu (A)$ is
a measure on the space of connections.

The motivation for the loop transform mirrors that of the use of the
fourier transform in ordinary relativistic quantum theory.  For one 
thing, for simple, non-intersecting differentiable loops, the
Wilson loops $T[\Gamma , A]$ are exact solutions of the Hamiltonian
constraints. Hence they play the role of plane wave states, in the
sense that the loop transform takes components in a basis in which
the equations of motion are easily represented.  Indeed, the dynamics,
whether formulated as a hamiltonian or a hamiltonian constraint,
acts at intersections, hence non-intersecting loops are in a sense
``free solutions''.    Furthermore
the loop states are invariant under $SO(3)$ gauge transformations so 
the Gauss's law constraint is automatically incorporated in the 
transform to the new representation. Finally, as we shall see, the
diffeomorphism constraint can be solved in the loop representation,
in terms of an infinite dimensional space of solutions, which is not
the case in the connection representation. 

The spin network states are particular combinations of loop states
which are linearly independent and which provide an orthonormal basis
for $H^{kinematical}$.  Basically they are combinations of loops
in which, whenever a number of loops coincide, the products of
traces in the fundamental, spin $1/2$ representation are decomposed
into sums of irreducible representations\cite{sn-roger,sn1}. 

An {\it abstract spin network} is an abstract
graph consisting of nodes connected by edges, which is labeled 
as follows:

Edges are labeled by spins, $j$, which are the irreducible 
representations of $SU(2)$.

Nodes are labeled by invariants (also called intertwiners or 
channels.) If a node has $n$ edges, with labels $j_1,...,j_n$
the invariant must be a map, 
$\mu : j_1 \otimes ....\otimes j_n \rightarrow Id$.  For each
set of representations  $j_1,...,j_n$ is a finite dimensional linear
space of such maps, called $V_{j_1,...,j_n}$, the space of
intertwiners. So the assignment of an invariant to each node is a 
choice of a vector in this space. 

An {\it embedded spin network} $\tilde{\Gamma}$  is an embedding of the 
abstract labeled graph $\Gamma$ into the spatial manifold $\Sigma$.  Any 
embedding is allowed so long as there are no intersections in the 
image not in the graph.  Given an embedded spin network
one gets a state $\Psi_{\tilde{\Gamma}}(A)$ in the connection 
representation by associating to each edge, $e$, labeled
with spin $j_e$ $e$ the parallel
transport of $A$ along that edge, in the representation $j_e$.
These are then tied together at each node, $n$ by the invariant
$\mu_n$ that labels it. The result is a gauge invariant function
of the connection, $\Psi_{\tilde{\Gamma}}(A)$, called a spin network 
state.

The measure $d\mu (A)$ has the property that the spin network states
provide an orthonormal basis in which the states $\Psi (A)$ can
be expanded. Thus, just like the fourier transform, there is an
infinite dimensional Hilbert space ${\cal H}$ of functionals
which has an inner product given by
\f
< \Psi | \chi > = \int d\mu (A) \bar{\Psi}(A) \chi (A).
\ff

This is called the kinematical inner product.

In the Euclidean theory the connections are real and the
inner product and loop transform are defined by an ordinary integral.
In the Lorentzian theory the connections are complex, and the
loop transform and inner product are defined by a contour
integral. The contour is usually taken to be the restriction to real
$SU(2)$ values of the connection. 
 
A spin network state can be expanded in terms of loop states, by 
expanding each representation in terms of symmetric products of fundamental 
representations. This is, however, rarely the best way to do a 
calculation.  Instead, it is usually best to make use of identities 
from representation theory.  

Since the spin network states make an orthonormal basis, 
one can then define the transform directly in terms of them by
\f
\tilde{\Psi} (\Gamma ) = \int d\mu (A) T[\Gamma , A]  \Psi (A) 
\ff
where $\Gamma$ here, and for the remainder of this paper 
labels an embedded spin network. In both the
connection and loop representation it can be shown that the spin 
network states are linearly independent and are an orthonormal basis
of $H^{kinematical}$\footnote{Spin network states were introduced 
in quantum gravity in \cite{sn1}.  But they were known already 
in the context of lattice gauge theory\cite{lattice-sn} (although not by that 
name.)  Much earlier Penrose\cite{sn-roger} introduced abstract spin 
networks and
argued that they would ultimately provide a discrete theory of quantum
geometry, as indeed they have been found to do.}. 

These basis elements are labeled $|\Gamma >$. A state in the loop
representation is then a functional of spin networks and is labeled,
\f
\Psi (\Gamma )= <\Gamma |\Psi >
\ff

We then have
\f
<\Gamma |\Gamma^\prime > = \delta_{\Gamma \Gamma^\prime}
\label{kip}
\ff

That is the inner product is zero unless the two embedded spin 
networks are identical, in which case it is equal to one. This is 
clearly very different from the Fock measure.  This one should think 
is good, given the result described in section ?? that there are no
exact states which correspond to linearized short wavelength 
gravitons. At the same time the reader may object that there is a 
problem, because $H^{kinematical}$ with the inner product
(\ref{kip}) is not separable.  This is a problem and it is a reason
why the loop transform is not useful when applied directly to
continuum formulations of gauge theories which are not diffeomorphism
invariant. (Although the technique is very useful in the context
of lattice gauge theory, where it has been used for a long time,
under different names\cite{latticeloop}.
This problem is however remedied when we go to 
the subspace of diffeomorphism invariant states, as we shall now see.

\subsection{Diffeomorphism invariant states}

Once we have transformed to the loop representation, we can study the 
action of both the diffeomorphism and hamiltonian constraints. The 
Hamiltonian constraint acts at vertices and generates three or four 
additional vertices.  As this is not the 
subject of the present paper we refer the reader 
to \cite{ham,roumen-ham,thomas-ham} where the 
action of the regulated constraint, and infinite dimensional spaces of 
solutions, are described in detail for $\Lambda =0$.  

The diffeomorphism constraint can be defined and solved exactly, 
because the kinematical loop representation carries an exact
unitary representation of the diffeomorphism group, defined for
a diffeomorphism $\phi$, by 
\f
\hat{U}(\phi ) \cdot |\Gamma > = |\phi^{-1} \cdot \Gamma >
\ff
It is easy to check that this representation is unitary under
the inner product (\ref{kip}).  A diffeomorphism invariant state
$ |\Psi >$
is then defined by $\hat{U}(\phi ) \cdot |\Psi > = |\Psi >$,
which implies that
\f
\Psi (\Gamma ) = \Psi (\phi^{-1} \cdot \Gamma )
\ff

It is easy to solve this and write an infinite number of 
diffeomorphism invariant states\cite{loop1}. 
For example, if $K(\Gamma )$ is
any invariant of knots, links and graphs, then
\f
\Psi_K (\Gamma ) =K(\Gamma )
\ff
is an exact diffeomorphism invariant state.  The space of 
diffeomorphism invariant states is denoted $H^{diff}$ may then
be constructed. It has an orthonormal basis labeled by the
diffeomorphism classes of the embedded spin networks\cite{}.  If
$\{ \Gamma \}$ denotes the diffeomorphism equivalence class of
the spin network $\Gamma$ then the elements of this basis are given by
\f
\Psi_{  \{ \Gamma \} } [\Gamma^\prime ] = 1 
\mbox{ if} \ \ \Gamma^\prime \in \{ \Gamma \} \mbox{and } 
0 \ \ \mbox{otherwise}. 
\ff

It is not difficult to show that this is a countable 
basis\footnote{This is trivial for graphs with vertices of valence $4$ or less, 
because then the diffeomorphism equivalence classes are countable. For 
higher valence the argument is more difficult because there are 
continuous parameters in the labels of the diffeomorphism equivalence 
classes. The spaces of these parameters are still finite dimensional, 
so it is still possible to pick a separable inner product. This is 
discussed in more detail in \cite{carlo-subtle}.}

For the case of $\Lambda=0$ an exact, rigorous formulation of the
loop transform has been constructed, in terms of a rigorous measure
on the space of connections, called the Ashtekar-Lewandowski 
measure\cite{}. Unfortunately this is not relevant for the case of
$\Lambda >0$, as we shall see. However there is another set of 
mathematical results that can be used to define the integral
transform in this case, coming from studies of Chern-Simons theory.

\subsection{The loop transform of the Kodama state}

The forgoing discussion of the loop transform and loop
representation has been only for $\Lambda =0$. 
We now consider the loop transform for $\Lambda \neq 0$.
The key point is that for $\Lambda >0$ we want the state space
to include the Kodama state, as we have shown it has all the
properties we would require of a vacuum or ground 
state\footnote{The loop transform for $\Lambda \neq 0$ was discussed
in \cite{sethme,sethroumenme,linking} and put into the form
presented here in \cite{tubes}.}.
The loop transform of the Kodama state is
\f
\tilde{\Psi}_K (\Gamma ) = \int d\mu (A) T[\Gamma , A]  
 {\cal N} e^{{\imath \kappa \over 4\pi} \int Y_{CS}}
 \label{kodamatransform}
\ff
where 
\f
\epsilon \kappa = k = {6\pi \over G \Lambda}.
\ff

Up to the possible factor of $\imath$ in $\kappa$,  this
is an expression that is well understood.   Indeed, 
for the Euclidean case, in which $\epsilon=1$, (\ref{kodamatransform}) 
is nothing but the expectation value, in Chern-Simons theory, of
an observable which is the Wilson loop associated to the embedded 
spin network $\tilde{\Gamma}$.
As first shown by Witten's \cite{witten-cs} 
$\tilde{\Psi}_K (\Gamma )$
is a known invariant of knots, links and graphs, which is the
Kauffman bracket\cite{lou-bracket}. 

The resulting functionals $\tilde{\Psi}_K (\Gamma )$ can be expressed
explicitly as a function of
$q=e^{{2\pi \imath \over \kappa +2}}$. Thus, for the Lorentzian
case the state can be gotten by continuing $q$
to imaginary values of $\kappa$.  
 
There is, however, a subtlety in the evaluation of
(\ref{kodamatransform}), which makes the loop representation
somewhat different for $\Lambda \neq 0$. This is that the 
integrals over the loop are singular, due to the presence
of the Chern-Simons term\cite{mecs}.  

As a result, the definition of the loop transform of the Kodama state 
requires a 
regularization procedure. There are several ways
to carry out the regularization, but however it is done the result is to 
introduce additional structure in
the definition of the Wilson loops. This structure is called framing,
and we must now review the basic ideas 
involved\cite{witten-cs,review-cs,mecs,louis-2d3d}. 

The problem is easy to see in perturbation theory. There the path 
integral must be supplemented with a gauge fixing condition, 
such as Lorentz gauge, $\partial_\mu A_\nu^i g^{\mu \nu}=0$. This
breaks not only gauge invariance but diffeomorphism invariance, 
because it requires that we introduce a background metric $g_{\mu\nu}$.
The choice of background metric is arbitrary and, as part of a gauge 
fixing procedure, it must not come into any physical quantities. Once
the gauge fixing has been introduced the linearized action may be 
inverted to find the propagator,  $
D_{\mu \nu}^{ij} (x,y)  \approx 1/|x-y|^{2}$
However then the leading order diagrams in $<T[\gamma , A]> $ 
involve a double integral over the loop, of the form
\f
<T[\gamma , A]> \approx
e^{\int_\gamma ds \int_\gamma dt \dot{\gamma}^\mu (s) \dot{\gamma}^\nu 
(t)D_{\mu \nu }(\gamma (s) ,\gamma (t)) }   + ...
\label{bad}
\ff
This expression has a singularity when $\gamma^\mu (s) = \gamma^\nu 
(t)$.

The result is that the expression (\ref{kodamatransform}) must be 
regulated. This must be done in such a way that diffeomorphism 
invariance and gauge invariance are not broken at the end of the 
calculation.  The known ways to accomplish this involve making the 
loop a two dimensional object, by expanding it in a direction 
linearly independent of $\dot{\gamma}^\mu (s)$.  One way to do this is
to attach a small vector $r^\mu (s)$ to the loop at each point such 
that $r^{[ \mu} (s) \dot{\gamma}^{\nu ] } (s) \neq 0$.  One can
then define a family of new loops by 
$\gamma_\epsilon^\mu (s) = \gamma^\mu (s) + \epsilon r^\mu (s)$ 
Then (\ref{bad}) is replaced by,
\f
<T[\gamma , A]> \approx \lim_{\epsilon \rightarrow 0}
e^{\int_\gamma ds \int_{\gamma_\epsilon} dt \dot{\gamma}^\mu (s) 
\dot{\gamma}^\nu_\epsilon 
(t) D_{\mu \nu }(\gamma (s) ,\gamma_\epsilon (t)) }   + ...
\label{good}
\ff
This is finite and well defined. However it turns out to depend on 
additional information, such as the Gauss linking number of $ \gamma$
and $\gamma_\epsilon$ for small $\epsilon$. This is diffeomorphism
invariant, but it is also additional information that must be
specified beyond the embedding (up to diffeomorphisms) 
of the loop $\gamma$ in the manifold $\Sigma$. This additional
information goes into the definition of a {\it framed loop}.  A framed
loop can be pictured as either a ribbon or a tube embedded in the
manifold\cite{louis-2d3d}.   

It turns out there are better ways to define framed loops and graphs, 
which relies on a connection to conformal field theory, which was the
basis of Witten's original paper \cite{witten-cs}. 

I will not go into the details of this construction here, but here
is a heuristic way to think about it\cite{louis-2d3d}.  
Let us consider first
a single loop. Imagine that the loop is blown up to a tube, 
introducing a boundary to the manifold $\Sigma$ with the topology
of a torus.  The torus may be considered ruled, which means that we can
identify, up to homotopy, the original loop in it. This is
sometimes called a tube ``with racing stripes.''

The 
connection on the loop can then be extended to a flat connection
on the two dimensional surface of the tube, restricted so that the 
traced
holonomy around any cycle which was also a cycle in the loop has
the same value as before, while the holonomy around any cycle
which was created by blowing up the loop to a tube is the identity
in the group. 

The nodes which connected the edges are now blown up to two
spheres. The points where the edges attached to the nodes
are blown up to circles, which are called punctures. 
The result is that a closed graph is blown up to a 
closed $2$-surface of some genus. However the $2$-surface
is ruled, which is to say there is a preferred, up to homotopy,
image of the original graph in the two surface. 

The connection is extended
to a flat connection on the $2$-surface, with the requirement that,
up to homotopy, any holonomy on a path which is in the image
of the original loop is equal to what it was in the original loop.

The analogues of the spin network states will be an orthonormal
basis, in some measure, of functions of the  flat connections on the $2$-surface.
However, these have to satisfy some identities, due to the
dependence on the framing.  For example, by going back to the 
original integral (\ref{kodamatransform}), it can be shown
that there are phase factors whenever one of the tubes are
twisted by $2\pi$ or whenever a punctures on a two sphere
representing a node is carried in a circle around one or
more other punctures\cite{witten-cs,mecs,review-cs}.  
These phase factors depend on the
coupling constant $\kappa$ that sits in front of the Chern-Simons
state. 

In conformal field theory the functions on the space
of flat connections with these properties are called conformal
blocks.  One way to notate them is by labeling the tubes
by representations, not of the group $SU(2)$, as before, but
by representations of the quantum deformed algebra 
$su_{q}(2)$ \cite{KL}.  Here the label $q$ is given in terms of
$\kappa$, and hence the cosmological constant, by  $q=e^{2\pi \imath \over 
\kappa+2}$.

Similarly, the two surfaces, with
punctures removed, may be labeled by intertwiners, or invariants, in the
representation theory of $su_{q}(2)$.  The structure of the 
representation theory of $su_{q}(2)$ turns out to capture exactly
the identities required by the regularization of the loop
transform of the Chern-Simons state.

We note that the level $\kappa$,  is in this case an imaginary
integer, related to the cosmological constant by eq. by (\ref{level}).
We found in section 11 that $k$ must be integer.

It is interesting to note that for Euclidean quantum gravity,
there would be an $\imath$ in the exponential in the Kodama
state and $\kappa$ would be an integer. In this case
$q$ is a root of unity.  When the deformation parameter 
 $q=e^{2\pi \imath \over k+2}$ is a root of unity,
    with $k$ an integer, the 
    irreducible representations are labeled by half-integral spins, 
    $j$, except that there is 
    a maximum irreducible representation, given by spin $j_{max}=k-1$.
    
The combinatorics of quantum spin networks
can also be defined a priori, as has been done by 
Kauffman and Lins in \cite{KL}.  One then introduces the notion
of an {\it abstract quantum spin network}, which is an extension of the
notion of an abstract spin network defined as follows, given by a
deformation parameter $q$.  

\begin{itemize}
    
    \item{}Edges are now tubes, which are ruled as described above. 
    They are labeled by a representation 
    of $su_{q}(2)$.
    
    \item{}Nodes are now punctured spheres, where each puncture,  
    represented by a little circle on the sphere, is a site by which 
    a tube is attached. These are labeled by intertwiners of 
    $su_{Q}(2)$. These are also 
    called the conformal blocks, on the punctured sphere, of the 
    conformal field theory called the $su(2)$ WZW model.

    \item{}The whole framed graph may be thought of as a ruled two 
    dimensional surface of some genus. The states associated with the  
    surface may also be taken to live in the space of intertwiners, or 
    conformal blocks on that surface. See \cite{louis-2d3d}.  

\end{itemize}

An embedded quantum spin network is a quantum spin network
embedded in the spatial manifold $\Sigma$ up to diffeomorphisms
of $\Sigma.$.  

Thus, we conclude that to accommodate the Kodama state in the physical 
Hilbert space, 
the whole spin-network basis must be quantum 
deformed to level $\kappa$.  Thus, {\it $\Lambda$ deforms the 
structure of the Hilbert space of physical states.}

We now note that the loop transform of the Kodama state 
(\ref{kodamatransform})
is also diffeomorphism invariant. Thus, (\ref{kodamatransform}) defines
an invariant of framed graphs, or equivalently ruled two surfaces,
embedded in the spatial manifold $\Sigma$. This is called the
Kauffman invariant. It is an important invariant of knots and
three manifolds.

\subsection{Excitations of the Kodama state}

We have just defined the loop transform of a single state,
the Kodama state.  To summarize, in  the two representations, we found 

\f
    <\Gamma | \Psi_K> = Kauffman (\Gamma )  \rightleftharpoons
   <A | \Psi_K>= e^{{\imath \kappa \over 4\pi} \int Y_{CS}}
 \ff
here $\Gamma$ is a quantum spin network, which is a ruled
two surface as described above, labeled with the representation
theory of the quantum group, imbedded in $\Sigma$ up to diffeomorphisms.
 
However the theory contains more than one state. In particular, in 
section 9 we found evidence that there is a large space of solutions
to all the constraints which can be gotten by perturbing the Kodama
state and which, at least for long wavelength approximate linearized
graviton states. 
We would like to take the loop transform of these states, in order
to to express the small perturbations of the Kodama state in a
diffeomorphism and gauge invariant language. 

Since the cosmological constant is coded, through $\kappa$, in the identities
that define the quantum spin networks, we may conjecture that
a space of the excitations of the Kodama state  
exists that retains the structure of the quantum group
deformation at level $\kappa$.  Thus, we want to consider a general state
of the form, 
\f
\Psi= \sum_\Gamma  c(\Gamma ) |\Gamma > 
\label{qstates1}
\ff
where $\Gamma$, which labels the basis states, are quantum
spin networks, as defined above, imbedded up to diffeomorphisms
in $\Sigma$.   

There is a natural inner product that can be defined on the
space of states (\ref{qstates1}), which is defined as 
follows\cite{tubes}.
Let us consider a two surface $\Delta$ embedded up to 
diffeomorphisms in $\Sigma$. This two surface has a 
space of flat $SU(2)$ connections ${\cal V}_{\Delta}$.
This space is finite dimensional and compact. The functionals
on ${\cal V}_{\Delta}$  are called the  
conformal blocks. They may be labeled by the intertwiners
of the $su_{q}(2)$ algebra.  If we neglect the ruling, then there
is a natural inner product on  ${\cal V}_{\Delta}$ given by
the representation theory of $su_{q}(2)$. In this inner
product, different states associated with different rulings
of the two surface are not generally orthogonal.  Instead, they
are connected by identities which extend the identities of
the representation theory of $SU(2)$ which give the definition
of the $6j$ symbols. 

To denote the inner product, let us denote a quantum
spin network by the pair $\Gamma =(  {\Delta}, \phi )$,
where $\Delta$ is the embedding of a two surface up to
diffeomorphisms in $\Sigma$ and $\phi$ is a state in ${\cal V}_{\Delta}$.
Then we can define the inner product by 
\f
<\Gamma | \Gamma^{\prime}> = \delta_{\Delta \Delta^{\prime}}
<\phi | \phi^{\prime}>_{{\cal V}_{\Delta}}
\label{quantumbasis1}
\ff
Thus, two quantum spin network states are always orthogonal if
they live on two surfaces $\Delta$ and $\Delta^{\prime}$ of
different genus, or if their two embeddings are non-diffeomorphic.
However, if the two surfaces are diffeomorphic their
inner product is given by the natural inner product in ${\cal V}_{\Delta}$.

The normalizable $\Psi$ of the form (\ref{qstates1}) with
inner product (\ref{quantumbasis1}) span a Hilbert space
${\cal H}^{qdiffeo}$.  This is an infinite dimensional 
state space, which is an extension of the space of
diffeomorphism invariant spin network states. 

We may note that in this inner produce two quantum spin network states
which differ by an extension of the recoupling identity, extended
to the quantum group, represent the same state. This contracts
somewhat the space of states compared with the ordinary 
spin network states, where there is no such identity. At the
same time, the space of states is still infinite dimensional,
because two states associated with quantum spin networks with
surfaces $\Delta$ of different genus will be orthogonal. Since
the genus can be an arbitrary non-negative 
integer this means the Hilbert space is infinite dimensional. 

A small perturbation of the Kodama state will then have the
form,
\f
c(\Gamma ) = Kauffman 
(\Gamma ) + k^{-1} \delta c
\ff

We may note that this automatically resolves a difficulty
with the theory at $\Lambda =0$, having to do with the
dynamics generated by the Hamiltonian constraint.
Briefly, here is the essence of the problem\cite{problem1,problem2} 
and its solution\cite{tubes}.

In either the hamiltonian or path integral (spin foam) formulation
of loop quantum gravity one finds that the dynamics consists of local
changes in the spin networks.  An example of such a local move
is the replacement of one trivalent vertex by a triangle containing
three new vertices. This is called the $1 \rightarrow 3$ 
move\cite{F-foam}.
By hermiticity, the time reverse of this move, denoted 
$3 \rightarrow 1$ move must also be present.

In fact  the $1\rightarrow 3$
and $3\rightarrow 1$ moves are all that are generated by certain forms
of the Hamiltonian constraint\cite{ham,roumen-ham,thomas-ham}.  
These arise from
a point splitting regularization technique where the different
operators that make up the hamiltonian constraint are split apart
in the spatial hypersurface $\Sigma$.  

However, there is another kind of move that can be defined on
spin networks, which is called the $2\rightarrow 2 $ move. This acts
on a pair of nodes that share a common edge, swapping their inputs, 
as in the usual angular momentum recoupling identities.

The key point is that the theory must contain these
$2\rightarrow 2$ moves if general relativity is to be recovered
as the low energy limit.  One reason is that only when these moves are
included can the dynamics satisfy the property that any finite spin 
network can evolve to any other one in a finite number of moves. A 
dynamics generated only by the $1\rightarrow 3$ and $3 \rightarrow 1$
moves divides the Hilbert space of spin network states into an infinite
number of sectors that do not mix under the dynamics.  This is 
inconsistent
with what is expected from the Einstein equations, for it implies
that there are an infinite number of observables, or constants of
the motion, that measure simple, quasi local properties of the states,
but which commute with the Hamiltonian constraint.  This would, at 
best, imply
that the theory was integrable, which we certainly do  not expect to 
be the case.  But even more
simply, one can show that in the absence of $2\rightarrow 2$ moves
information is not propagated in the quantum spacetimes, because it 
can be shown that information at one node of an initial spin network
never propagates even to nearby nodes.

There are still other difficulties
with the forms of the Hamiltonian constraint
found in \cite{} that
neglect the $2 \rightarrow 2$ moves.  These involve problems with the 
algebra of constraints\cite{} as well as problems with preserving
positive energy in the quantum theory\cite{}.

The inclusion of the $2 \rightarrow 2$ moves solves many, if not
all, of these problems. It is also possible to see that these
moves are required by spacetime relativistic
invariance, and that the problem arises from the fact that the
regularization procedures used in the construction of the operators
impose a preferred spatial slicing in which the point splitting is done.

It is then very interesting to notice that the $2\rightarrow 2$ moves
are automatically included once the theory is expressed in terms
of quantum deformed spin networks. The reason is that when $q \neq 1$ 
the states that differ by these moves are no longer orthogonal to
each other in the inner product just defined\cite{tubes}.
Thus, any state has a non-vanishing amplitude to be found
in a state that differs by an application of one or more
$2\rightarrow 2$ moves.

\section{The inclusion of spacetime boundaries} 

Up till this point we discussed the quantum theory for
a spatial manifold with compact topology. We now consider
the case in which the region of spacetime which is
treated quantum mechanically has a boundary. 

In section 6 we discussed the general problem of how to
deal with boundaries in the classical theory, after which
we studied a class of boundary conditions called 
the ``Chern-Simons boundary
conditions'', eq. 
(\ref{CSboundarycondition}).    

These boundary conditions can be quantized in the Hamiltonian 
approach and the physical
picture that results is the following\footnote{The Chern-Simons
boundary conditions were introduced in \cite{linking} first for
the Euclidean case.  Kirill Krasnov realized they apply also
to the horizon of a black hole\cite{kirill1}, this led to the
development of the {\it isolated horizon} approach to the quantum
geometry of horizons\cite{isolated}.  The application of Chern-Simons
boundary conditions to time like boundaries with $\Lambda \neq 0$
was developed in \cite{me-holo} and applied, for $\Lambda <0$ to
supergravity in \cite{yime-holo}.}

Note that when eq. (\ref{CSboundarycondition}) is imposed, neither
the metric nor the connection is fixed on the spatial boundary.
So the boundary is, like $\Sigma$,  a manifold without metric.

The quantum spin networks can end on the spatial boundary. They 
end in punctures, labeled  by reps of $SU_q(2)$, which we
denote $\{ j\}$.

To understand the physics of the boundary we can use the fact that the
operator which measures the area of the boundary is well defined.
Its eigenvalues are associated with different sets of punctures,
and have the form,
\f
A[{\cal B} ] = {\hbar G \over 2}  \sum_{j}\sqrt{C_{j}}={ \hbar G 
\over 2}
\sum_{j}\sqrt{j(j+1)} + O(1/k)
\label{Barea}
\ff
where $C_{j}$ is the quadratic Casimer of the quantum group $su_{q}(2)$.

Thus, for an eigenstate of $A[{\cal B}]$
the boundary, $\cal B$, can be represented as a compact surface
with a fixed set of punctures. 

For each choice of punctures, the Hilbert space decomposes
into a product of a bulk piece and a boundary piece. Thus,
the whole hilbert space in the presence of the boundary
has the form, 

\f
H= \sum_{\{ j\} } H_{\{ j\} }^{bulk} \times H_{\{ j\} }^{boundary}
\ff

$ H_{\{ j\} }^{bulk}$ has a basis given by quantum spinnets 
that end on the punctures $\{ j \}$.  

To understand the boundary Hilbert space, note that 
the dual of the frame field, pulled back into the boundary,
$E_{ab}= \epsilon_{abc}E^{c}$ (which is also the self-dual two
form of the metric, pulled back into $\cal B$)
is fixed by (\ref{CSboundarycondition})
to be proportional to the curvature. Hence the only degree of
freedom which survives on the boundary are the components of the connection,
$A_{a}$, pulled back to the boundary.

The next thing to notice is that there is a time derivative in the boundary 
term, as it is after all the action of the three dimensional 
Chern-Simons theory.  The
Poisson brackets and hence the commutation relations are then
altered on the boundary. On the boundary the canonical commutation
relations from Chern-Simons theory hold\cite{linking},
\f
[ A_{a}(\sigma ) , A_{b} (\sigma^{\prime})]  = {2\pi \hbar \over k}
{\imath \over \epsilon}
\epsilon_{ab} \delta^{2} (\sigma , \sigma^{\prime} )
\ff
where we recall that $\epsilon= \imath$ for the Lorentzian
theory and $\epsilon =1$ for the Euclidean continuation.  
Thus, in either case the boundary Hilbert space is related to the Hilbert
space of Chern-Simons theory, with $\kappa =\epsilon k$.  
To see how we consider the
effect of the boundary condition. 

The action of the boundary condition (\ref{CSboundarycondition})
can best be understood by integrating it against a test
function $g(\sigma )$ on $\cal B$. They then have the form
\f
\int_{\cal B} d^2 \sigma^{ab} F_{ab} \ g (\sigma ) = {\Lambda \over 3} 
\int_{\cal B} d^2 \sigma^{ab} \epsilon_{abc} E^{c} \ g (\sigma ).
\ff
The operator $E^{c}(\sigma )$ will only be non-zero if
it acts at a point where an edge of the quantum spin network
meets the boundary. So we deduce that for all regions
in which there is no edge attached, the curvature
$F_{ab}$ vanishes. 

When there is an edge attached in the region where $g$ has
support the effect is the same as a delta function singularity
of the form,
\f
F_{ab}^{i} (\sigma ) \approx {2\pi   \over k} X^{i}_{j}
\epsilon_{ab} \delta^{2}(\sigma , \sigma^{\prime} )
\label{fixed}
\ff
where $j$ is the representation labeling the edge at a puncture
at the point $\sigma^{\prime}$ and $X^{i}_{j}$ is an element
of the lie algebra in the conjugacy class of $j$. 

Thus, for a fixed set of punctures, $\{ j \}$, the boundary
Hilbert space is the Hilbert space of Chern-Simons theory,
at level $\kappa$, on the two surface, which is the boundary with
the fixed set of punctures. 
This is called ${\cal V}_{\{ j \}}$.  
The boundary condition has just
fixed the punctures of the boundary to be points at which
ends of the quantum spin nets are attached. 

The Hilbert space of Chern-Simons theory on a punctured
surface, with (\ref{fixed}) imposed is well 
understood\cite{louis-2d3d,witten-cs,review-cs}
for the case of real level, which is the case of the Euclidean
continuation.
It has several characterizations. First, it is isomorphic
to the space of intertwiners of the quantum group, for
the products of the representations labeling the punctures. 
It is also the space of conformal blocks of the
$SU(2)$ Wess-Zumino-Witten theory on the punctured surface.

Given that the state we have identified as the ground
state of the theory satisfies the $KMS$ condition to
be a thermal state, it makes sense to seek to derive
thermodynamic relations concerning the quantum theory.
When we do this we should use the Euclidean signature version of the theory,
as it is the Euclidean histories that are integrated over in
the path integral representation of a thermal state. 
Thus, in what follows we take $\kappa = k = 6\pi / \Lambda$
to be an integer. 

\subsection{Automatic satisfaction of the Bekenstein bound}

The boundary hilbert space we have defined is finite dimensional
for a finite number of punctures.  Moreover the Bekenstein
bound\cite{bb} is automatically satisfied, as we now show\cite{linking}.

For a fixed set of punctures, $\{ j \} $ the Hilbert
space is an eigenspace of the operator that measures the
area of the boundary, with eigenvalue given by eq. (\ref{Barea}).
Thus, the eigenspaces of the area operator comprise the
finite dimensional vector spaces, ${\cal V}_{\{ j \}}$.
The exact value of the dimension of ${\cal V}_{\{ j \}}$ is
given by the Verlinde formula\cite{verlinde}. However
it is easy to estimate for a large number of punctures, and
large $k$. In this case we may approximate its dimension by the dimension
of the space of invariants in the product of the representations
$j$. For large $k$ the quantum dimensions of these
representations are close to the dimensions of the
representations of the classical lie algebra. 
Neglecting a small factor this gives
\f
dim {\cal V}_{\{ j \}} = \prod_{j} (2j+1)
\ff
The Bekenstein bound is the statement that
\f
\ln dim {\cal H}^{bound} < {A[{\cal B}] \over 4G_{R} \hbar} 
\ff
where $G_{R}$ is the renormalized value of Newton's constant,
which is the one measured in the classical theory.  This will be
related to the $G$ of the fundamental theory by
\f
G_{R} = c G
\ff
where $c$ is a multiplicative renormalization constant.  Since the
theory has an ultraviolet cutoff we expect $c$ to be finite,
in fact $c$ is of order unity, as we now see.

By comparing the expressions for area and the dimension of the 
boundary Hilbert space, we see that the Bekenstein bound will 
always be satisfied, for large $k$
and large area, so long as 
$c$ can be chosen such that it is always the case that
\f
\sum \ln (2j+1) < {c \over 4} \sum_{j} \sqrt{j(j+1)}
\ff
This is the case, by inspection, and we see that the
constant is fixed by the smallest representation, $j=1/2$.
So we find, $c= {ln(2) \over \sqrt{3}}$.  

So we learn that the Bekenstein bound is satisfied up to
a constant of order one, in the bare newton's constant.  If we
require that it hold exactly with the $1/4$ then we fix
the renormalization of the gravitational constant.

\subsection{Applications of the boundary theory}

As mentioned in section 6, there are several cases in which the
boundary theory is of this general form, depending on whether 
the spacetime boundary is null or timelike.  There is a large
literature about the case where the boundary is an horizon,
satisfying certain conditions, which amount physically to the
condition that the spacetime is stationary in the neighborhood
of the horizon.  Using the appropriate specialization of the
Chern-Simons boundary conditions the black hole entropy may be
derived and, in fact, explained, as the log of the number of
quantum states at the boundary which are compatible with the 
condition that the boundary is, locally, an horizon\cite{kirill1,isolated}.
There are further results which show that, at least heuristically,
the Hawking radiation can be derived\cite{kirill-radiate}. 
Most interestingly, there
are calculations that indicate that the Hawking entropy has a fine
structure and that there are corrections to the relationship
between area and entropy\cite{fineBH,kirill-radiate,logcorrect}.  

Another case that has been studied is that of time like boundary
conditions, for the case of negative $\Lambda$ \cite{me-holo,yime-holo}.  
There
are indications here that something like the AdS/CFT correspondence
may exist in $3+1$ dimensions, when the limit is taken in which
the area of the spatial boundary goes to infinity\cite{inprogress}.

There has been less work on the imposition of boundaries
when the cosmological constant is positive, but some of the
results of \cite{me-holo,yime-holo} apply to this case as well. I 
will describe here just the main features the theory,
details will appear elsewhere\cite{inprogress}.

It is important to stress that for the interpretation of
the constant $c$ as a finite renormalization of Newton's
constant to hold, it must be universal, and all calculations
must lead to a single value for $c$.  This is so far the case
in general relativity, whatever type of boundary or horizon
has been studied, $c$ turns out to be universal.

However, $c$ does differ in supergravity, where it depends
on the number of supersymmetry generators\cite{yime-holo}.
Hence if there were no other way to detect supersymmetry,
it could in principle be detected by measuring the quanta
of area in units of the observed, $G_{R}$.

\subsection{The N-bound}

The boundary theory developed here gives some insight 
into an important conjecture
of Banks, called the $N$ bound\cite{N-banks}.  
Motivated by the finiteness
of the entropy of a horizon in deSitter spacetime, Banks conjectured
that a quantum theory of gravity will have only a finite number, N, 
of degrees of freedom when $\lambda$ is positive, with
\f
N = {3\pi \over \Lambda }
\label{Nbound}
\ff
For a semiclassical quantum field theory in deSitter spacetime, 
the $N$ bound follows from
Bousso's form of the holographic bound\cite{N-bousso}. Bousso further
argues that the bound holds for any semiclassical theory in a
spacetime with $\Lambda >0$, at least so long as certain
kinds of matter fields are excluded\cite{N-bousso}.

Here is the basic semiclassical argument for the $N$ bound.
In deSitter spacetime the horizon of any observer
has area,
   \f
A_{max} = {12 \pi \over \Lambda}
\ff
This is hence the largest surface from which an observer in
deSitter spacetime could receive information.  If we believe
in the Bekenstein bound then we would say that the most
information that could be read by an observer on the horizon
is $N_{max}$ where
\f
N_{max} = {A_{max} \over 4} = {3 \pi \over \lambda}
\ff
This is hence the maximum dimension of a Hilbert space needed
to represent the information that an observer in deSitter space
could measure.

To generalize beyond deSitter spacetime, one may make use of results 
that suggest that, when $\Lambda >0$, 
any spacetime satisfying the usual energy conditions
will asymptotically approach deSitter spacetime.  This is because,
for the usual forms of matter, 
in time the exponential expansion dilutes the effect of matter and
gravitational radiation, so that the effective energy momentum tensor
is dominated by the cosmological term.

However, if the $N$ bound is really fundamental, it should
apply not to the degrees of freedom of a semiclassical
quantum field theory, in a fixed classical spacetime background,
but to the full quantum theory of gravity.  Thus, it is interesting
to see if the bound may be derived directly from the full quantum
theory of gravity in the case $\Lambda >0$.  

We consider the case that $k$ is
large, but finite, so that $\Lambda$ is small in Planck units.
We consider also the case that the universe is described by a
thermal state close to the Kodama state. In this case the universe
can be thought of as close to thermal equilibrium, which corresponds
to the fact that classically deSitter spacetime is stationary.  We can then
use the Euclidean formalism to describe what an observer
measures. 

Since $k$ is very large, the horizon is far from the observer and
the observer receives information from quanta which are
radiated from the horizon to the observer.

We next note that the gauge group $SO(3)$ corresponds 
to the freedom of the observer to rotate locally.
It then follows that {\it the information received by the observer 
at any one time should\footnote{This requirement brings to mind
the origin of the word {\it universe} as ``that which turns as one.''}
be contained in an irreducible representation of $SO(3)$. }  

However, we have seen that when $\Lambda \neq 0$ the
representation theory of the local rotation group is quantum
deformed. This means that there are non-trivial quantum effects,
due to the presence of the cosmological constant,  \
which modify what happens to the observer's view of the 
universe when he rotates.  So the preceding statement should
be modified to say that {\it the information received by an 
observer at any one time should be contained
in an irreducible representation of $SU_{q}(2)$.}

However, when $k$ is a finite integer 
there is a largest representation of $SU_{q}(2)$.  As a result there 
is a limitation on the amount of
information that could be measured by a local observer in
at any given time.
The bound on the information measurable corresponds to the dimension of the 
largest irreducible  representation of $SU_{q}(2)$.

Now, the largest representation with $k$ finite  is\cite{KL}
\f
j_{max}={k \over 2}= {3\pi \over \lambda}
\ff

But $j_{max}$ is also the number of degrees of freedom.
This is because, for 
a given area, the dimension of the space of intertwiners, or 
invariants,
is extremized when the spins have the lowest possible 
value.
Thus, maximal entropy screens have some number, $n$, of spin $1/2$ punctures.
The entropy  such a  screen is then given for large $n$ 
and large $k$ approximately by 
\f
S(n) \approx  \ln (dim [ {\cal H}_{1/2} ])  n  =n \ln (2).
\ff
The corresponding number of degrees of freedom is
\f
N = {S(n) \over \ln(2)} = n
\ff
because entropy is counted in units of Shannon information.

But $ N_{max} = {k \over 2} $, so we have
 
\f
N_{max} = {3 \pi \over \Lambda }
\ff
This is the $N$-bound.

To summarize, we have given a physical argument in support of a
connection between causality, measurement and irreducibility,
in the context of loop quantum gravity.  The connection is that
the information received by any local observer at any one time
must be contained in some irreducible representation of the
local gauge group. Otherwise, there
are quantum states, observable by a local observer in spacetime,
that do not transform irreducibly when that observer rotates locally.
We have then shown that this conjecture implies
the $N$-bound.  

\subsection{How to do quantum cosmology with horizons}

Before closing, I want to make a few remarks concerning some
of the deeper problems the quantum theory of gravity faces,
concerning the application of the quantum theory to cosmological
spacetimes in which the observers, along with their measuring
instruments and clocks, must be considered part of the quantum
system.  While this problem remains open, I would like to 
describe here a new approach which has been proposed and
developed by people working in loop quantum gravity, which has
come to be called {\it relational quantum cosmology.}

The problem of quantum cosmology is especially acute
for the case
$\Lambda >0$ as classical spacetimes 
 with $\Lambda >0$ generically have horizons. This 
means that no real observer inside the universe can observe the
whole universe.    In such circumstances there is a limit to much of the 
universe any observer can see, no matter how long they wait. 

To proceed we may distinguish three closely related issues:

-What is the right formal structure for quantum theory in a 
cosmological context? Should it be based on the conventional Hilbert 
space and algebra of observables, or on something more exotic?

-What is the right measurement theory for quantum cosmology? How do 
we deduce predictions for real experiments from the mathematical 
states and operators, or whatever replaces 
them? Here it must be stressed that if the theory is going to 
be compared with observations we can make, the measurement theory 
cannot just produce predictions for asymptotic observers, at very late 
times. To be useful for us the measurement theory 
must make sense of observers inside the universe, at times 
which, however large they may be on Planck scales, are still short 
compared to the lifetime of the universe, or the time it takes the 
universe to reach some final asymptotic state. 

Thus, whether or not there are ultimately horizons, it is still the 
case that finite observers such as ourselves only have access to a 
small fraction of the whole universe. Thus, whatever the value of the 
cosmological constant the measurement theory must allow a situation 
where the only physically possible observations are partial 
observations. 

-However, there may still be special features of a quantum cosmology 
with a positive cosmological constant, which the measurement theory 
must incorporate.

Over the last fifty or so years there have been several serious 
attempts to solve the problem of quantum cosmology.   These attempts 
are of two kinds. There are approaches to quantum cosmology that take 
the mathematical structure of the theory to be the same as that of 
ordinary quantum mechanics which we may call ``conventional quantum 
cosmology''. All the other approaches propose that the formal structure 
of quantum theory must be modified for the theory to be sensibly 
applied to cosmology.

Quantum cosmology is a controversial subject.

The problem is, briefly, that the major interpretations of quantum 
mechanics rely on a division of the world into a classical part, which 
contains the observer, her clocks and measuring instruments and a 
quantum part, which includes the system under study.

The Copenhagen interpretations of quantum theory, as propounded, in 
somewhat different forms by Bohr, Heisenberg, Dirac, von Neumann and 
others all explicitly introduce this split. Since then other 
interpretations of quantum mechanics have been proposed, such as the 
Everett interpretation, the many universes interpretation, the 
consistent histories interpretation, etc, that do not rely on a split 
of the world into a classical and a quantum part.  Thus, these 
interpretations seem to allow the whole universe to be described as a 
single quantum system.  However, to make a very long story short, 
when developed in practice, each of these interpretations relies on 
an additional ad hoc hypothesis, which is introduced to make a 
connection between calculations in quantum theory and real 
observations made by real observers.  While this is not the subject 
of this paper, I believe it is the case that none of these 
interpretations really solved the problem of how to apply quantum 
theory in a cosmological situation.  

As a result, over the last several years a new approach has
been proposed\cite{louis-holo}-\cite{weakholo}.  
The main idea of these approaches is to take seriously
Bohr's idea that the measurement theory of any quantum theory must
be based on a split of the universe into two parts, the quantum
system, which is modeled in a hilbert space, and a classical part,
which includes the observer along with her clocks and measuring
instruments.  Bohr always stressed that the division between these
two worlds must be arbitrary, so that the physical predictions of the
theory do not depend on how it is made.  Bohr also stressed that the
quantum state was not so much an objective property of the quantum
system as it was something dependent on both sides of the split, as it
contained the information needed to deduce what will be the result of
all of the various measurements accessible to the classical observer,
carried out on the quantum system.  

These ideas of Bohr are among the most mysterious features of his 
thought, for it appears to imply that the quantum state, and the
Hilbert space in which it lives, are not completely objective
properties of the quantum system, as they depend to some extent
on how the quantum/classical split is made. 

The main idea of relational quantum
cosmology is to take Bohr seriously, by following the methodology
of special relativity. Rather than looking for a preferred observer,
or a preferred split of the world into its quantum and classical parts,
the idea is to include in the formalism of the quantum theory all 
possible ways of dividing the universe into two parts.  There are then
many Hilbert spaces, one for each possible way to divide the universe
into a quantum and classical part. Each Hilbert space becomes an 
objective property, but not of the system, but of  
the boundary between the classical
and quantum world.

The basic ideas of relational quantum cosmology  may be summarized as follows:

\begin{itemize}
    
\item{}Crane: {\it  Hilbert spaces are associated with boundaries that 
split the universe into parts.  By the relationship of GR to TQFT 
these will be described in terms of finite dimensional state spaces. 
Hence the Bekenstein bound.}\cite{louis-holo}

\item{}Rovelli:  {\it  Each hilbert space describes the information one 
part of the universe has about another part.  The various measurement 
problems are solved by paying attention to the fact that different
observers record their measurements in different Hilbert spaces. }
\cite{carlo-relational}

  \item{}Dyson: {\it For each observer, the description of the past is 
    classical, while the present and future are described quantum 
    mechanically.}\cite{Dyson}
    
    \item{}Markopoulou: {\it The natural way to divide the world into
    quantum quantum systems and observers is to use the causal 
    structure of spacetime.  In a cosmology there is then a Hilbert
    space for each local region of spacetime, in which is represented
    the information that arrives there from its causal past. These
    are tied together by maps which incorporate the dynamics and
    express the causal relations}\footnote{These structures are 
    mirrored also in 
    the algebra of observables of any classical, relativistic
    cosmological theory which  is not Boolean but Heyting. 
    This is due to the  the fact that no observer can give
  truth values to all propositions about the history of the universe.
    This algebra has no representation in terms of projection operators in a 
    single hilbert space.}.\cite{Fotini-QCH}

\item{}Butterfield and Isham: {\it The right mathematics for relational 
quantum theory is topos theory and more particularly, the structure
of the Hilbert spaces for the quantum theory of spacetime is that
of a presheaf over a partially ordered set. }\cite{BI}

\end{itemize}

Reduced to a slogan, relational quantum cosmology maintains that, 
``Many quantum states to describe one universe, not one state 
describing 
many universes."

From this brief description, it is clear that relational quantum
theory is closely related to the Holographic Principle, according
to which a quantum gravitational system may be described in terms of 
information measurable on its boundary. Indeed, the earliest, to my
knowledge, statement of the holographic principle, is in a paper by
Louis Crane in which he formulated also the basic idea of
relational quantum theory.  That paper was, in turn, inspired
by the relationship between quantum gravity and topological quantum
field theory.
 
A version of the holographic principle compatible with the
ideas of relational quantum cosmology was then proposed 
in \cite{weakholo}\footnote{Its relationship with other proposed
form of the holographic principle is discussed in \cite{weakstrong}.}.
it is called the {\it weak holographic principle} and may
be summarized as follows,

\begin{itemize}

    \item{}A quantum spacetime has a discrete causal structure,
    in which the events are transitions, or local moves, 
    in a (quantum) spin network.  The analogue of spacelike
    slices are (quantum) spin networks\footnote{The whole
    $3+1$ dimensional structure is called a causal spin foam, and
    is described in \cite{F-foam,tubes,stume}. }.
    
\item{}The causal structure extends to spacelike surfaces,
which are defined in terms of the topological dual of the quantum
spin networks. Thus, to each spacelike two surface is associated
a set of representations $\{ j\}$ which are the labels of the
edges it is dual to, or intersects.

\item{}Any spacelike  surface in a quantum spacetime can be
considered a channel through which quantum 
information flows from its causal past to its causal future.  
To each such spacelike two surface is associated the finite
dimensional Hilbert space ${\cal V}_{\{ j \}}$.  The information
flowing through the surface is represented by an operator
in  ${\cal V}_{\{ j \}}$.  

\item{}All measurements are made on such surfaces, hence all
observables in a quantum spacetime are operators in 
some ${\cal V}_{\{ j \}}$.

\item{}The area of the surface is another name for its capacity as a 
channel of quantum information.   Thus, the Bekenstein bound, which
as we have shown is automatically satisfied, is 
interpreted as a reduction of area to a measure of information flow.

\end{itemize}
 
If these ideas are right, they should apply to quantum gravity
with a cosmological constant.  Each observer in deSitter spacetime
is able to receive information from only that part of the spacetime
interior to her horizon.  We indeed found that we could develop
the semiclassical theory in the interior of each such region,
characterized by a flat slicing of a region of deSitter spacetime.
In the full quantum theory the horizon may be represented by a 
boundary, on which the self-dual
boundary conditions discussed here may be applied. 

\section{Conclusions and further developments}

In this paper we have presented a number of results, some old and some
new, which together support a claim that for $\lambda >0$ 
and in $3+1$ dimensions loop quantum
gravity gives a satisfactory quantum theory of gravity.
The main results were stated in the introduction.

There are a number of related developments which I did not have
space to mention here.  These include

\begin{itemize}
    
    \item{}The idea that gravity is a constrained topological field
    theory turns out to be very powerful when applied to the derivation
    of measures for the path integral formulations of quantum gravity,
    which are called spin foam models\cite{mike-foam,BC,baezfoam}.

\item{}A related model, called a causal spin foam model, has
been developed for $\Lambda >0$\cite{tubes}.

\item{}The Chern-Simons state has been studied in the context of 
reduced quantum cosmological models, in which only a few degrees
of freedom are retained, such as in the quantum Bianchi 
models\cite{qc-cs}. 

\item{}The Chern-Simons state can be transformed to the triad, or 
frame field representation\cite{triad}.
    
    \item{}Soo has described
    an expansion around the Kodama state 
    in powers of $\lambda$ \cite{chopin-new}.

    \item{}A strong coupling expansion, in powers of $1 \over \lambda$
    was proposed in \cite{ham} and has been developed in some detail in
    \cite{GP-lambda}.
    
    \item{}The Kodama state can be used to generate an infinite number 
    of physical states at $\lambda =0$\cite{GP-generate}.
    
    \item{}There are related results also for $\Lambda <0$, including
    a number that extend to supergravity for 
    $N=1$ and $N=2$ \cite{yime-holo,yi-super,super}.

\end{itemize}

There remain several things that must be done to clinch the
case. These include,  1) development of the perturbation theory
for the excitations of the Kodama state, leading to computations of
corrections to graviton-graviton and graviton-matter scattering.  
This will require an understanding of how to expand small excitations 
of the Chern-Simons state in the basis of exactly gauge and 
diffeomorphism invariant states. 
2)  More study of the corrections to the energy momentum 
relations\footnote{This is in progress with Carlo Rovelli and
additional results will be reported in 
\cite{newcarlolee}.}, including calculations for photons and
protons and an understanding of the consequences
for either breaking or modifications of  lorentz invariance at Planck
scales.  3)  Development of the relational framework for quantum
cosmology, to give a complete theory of quantum spacetimes with
$\Lambda >0$.

\section*{ACKNOWLEDGEMENTS}

I wish first to thank my collaborators on different aspects
of the problem of quantum gravity with a cosmological constant,  
 Roumen Borissov, Yi Ling, Joao Magueijo, Seth Major, Fotini Markopoulou, 
Carlo Rovelli and Chopin Soo.   I also
have learned a lot about this and related subjects from discussions 
and correspondence with
Giovanni Amelino Camelia, John Baez, Tom Banks, Raphael Bousso, 
Louis Crane, Laurent Freidel, Lou Kauffman, Renate Loll, Michael Reisenberger and
Artem Starodubstev.
I am especially grateful to Laurent Freidel and Chopin Soo for 
comments which greatly improved the presentation. 
This work was
supported generously by The National Science Foundation and the Jesse Phillips
Foundation.


\begin{thebibliography}{99}

\bibitem{carlo-review}C. Rovelli,  Living Rev. Rel. 1 (1998) 1, gr-qc/9710008.

\bibitem{loopreviews}R. Gambini and J. Pullin, Loops, knots, gauge theories and 
quantum gravity Cambridge University Press, 1996.

\bibitem{spain}L. Smolin: in 
Quantum Gravity and Cosmology, eds J Perez-Mercader et al, World
Scientific, Singapore 1992; The future of spin networks gr-qc/9702030 
in the Penrose
Festschrift.

\bibitem{sen}A. Sen, On the existence of neutrino zero modes in vacuum 
spacetime J. Math. Phys. 22 (1981) 1781, 
Gravity as a spin system Phys. Lett. B11 (1982) 89.

\bibitem{abhay}Abhay Ashtekar, 
New variables for classical and quantum gravity," Phys. Rev. Lett.
57(18), 2244-2247 (1986).


\bibitem{kodama}H. Kodama, Prog. Theor. Phys. 80, 1024(1988);
Phys. Rev. D42(1990)2548.

\bibitem{chopinlee}L. Smolin and C. Soo, The 
Chern-Simons Invariant as the Natural Time Variable for
Classical and Quantum Cosmology, Nucl. Phys. B449 (1995) 289, gr-qc/9405015.

\bibitem{chopin-new}C. Soo, Wave function of the Universe
and Chern-Simons Perturbation Theory, gr-qc/0109046.

\bibitem{GAC1}G. Amelino-Camelia et al, Int.J.Mod.Phys.A12:607-624,1997;
G. Amelino-Camelia et al Nature 393:763-765,1998;
J. Ellis et al, Astrophys.J.535:139-151,2000;
J. Ellis, N.E. Mavromatos and D. Nanopoulos,
Phys.Rev.D63:124025,2001; ibidem astro-ph/0108295.

\bibitem{AC-Piron} G. Amelino-Camelia and T. Piran,
Phys.Rev. D64 (2001) 036005.

\bibitem{seth}Tomasz J. Konopka, Seth A. Major,  
``Observational Limits on Quantum Geometry Effects'', New J.Phys. 4 (2002) 57.
hep-ph/0201184;  Ted Jacobson, Stefano Liberati, David Mattingly,
`` TeV Astrophysics Constraints on Planck Scale Lorentz Violation'',
hep-ph/0112207.

\bibitem{testreviews}Subir Sarkar, `` Possible astrophysical probes of 
quantum gravity'',  Mod.Phys.Lett. A17 (2002) 1025-1036, 
gr-qc/0204092.

\bibitem{linking}L. Smolin, Linking topological quantum 
field theory and nonperturbative quantum gravity,
J. Math. Phys. 36(1995)6417, gr-qc/9505028.

\bibitem{kirill1}K. Krasnov, On Quantum Statistical 
Mechanics of a Schwarzschild Black Hole , grqc/9605047, Gen. Rel. Grav. 
30 (1998) 53-68; C. Rovelli, ``Black hole entropy from loop quantum gravity," grqc/
9603063.


\bibitem{isolated}A. Ashtekar, J. Baez, K. Krasnov, Quantum Geometry of 
Isolated Horizons and Black
Hole Entropy gr-qc/0005126; A. Ashtekar, J. Baez, A. Corichi, K. Krasnov, 
``Quantum geometry and black hole entropy," gr-qc/9710007,
Phys.Rev.Lett. 80 (1998) 904-907.

\bibitem{me-holo}L. Smolin, A holograpic formulation of quantum 
general relativity, Phys. Rev. D61
(2000) 084007, hep-th/9808191.

\bibitem{yime-holo}Y. Ling and L. Smolin,
Supersymmetric spin networks and quantum supergravity,
Phys. Rev. D61, 044008(2000), hep-th/9904016;
Holographic Formulation of Quantum Supergravity, hep-th/0009018,
Phys.Rev. D63 (2001) 064010.

\bibitem{N-banks}T.Banks, ``Cosmological Breaking of Supersymmetry?'',
hep-th/0007146.

\bibitem{N-bousso}Raphael Bousso, ``Positive vacuum energy and the N-bound'',
hep-th/0010252,  JHEP 0011 (2000) 038.

\bibitem{tedlee}T. Jacobson and L. Smolin, Nucl. Phys. B 299 (1988) 295.

\bibitem{loop1}C. Rovelli and L. Smolin, Knot theory and quantum theory, Phys. Rev. Lett
61(1988)1155; Loop representation of quantum general relativity, Nucl. Phys.
B331(1990)80-152.

\bibitem{loop2}R. Gambini and A. Trias, Phys. Rev. D23 (1981) 553, Lett. al Nuovo Cimento
38 (1983) 497; Phys. Rev. Lett. 53 (1984) 2359; Nucl. Phys. B278
(1986) 436; R. Gambini, L. Leal and A. Trias, Phys. Rev. D39 (1989) 3127.

\bibitem{sn-roger}R. Penrose, Theory of quantized directions unpublished manuscript;
in Quantum theory and beyond ed T Bastin, Cambridge U Press 
1971, in 
Advances in Twistor Theory, ed. L. P. Hughston and R. S. Ward, (Pitman,1979) p. 
301in Combinatorial 
Mathematics and its Application (ed. D. J. A. Welsh) (Academic Press,1971).

\bibitem{sn1}C. Rovelli and L. Smolin, Spin networks and quantum gravity, 
gr-qc/9505006, Physical
Review D 52 (1995) 5743-5759.

\bibitem{volume}C. Rovelli and L. Smolin Discreteness of area and volume in 
quantum gravity, Nuclear
Physics B 442 (1995) 593. Erratum: Nucl. Phys. B 456 (1995) 734.

\bibitem{renate-volume}R. Loll, ``The volume operator in discretized quantum 
gravity," Phys. Rev.

\bibitem{volume2}Roberto De Pietri and Carlo Rovelli, Geometry eigenvalues 
and the scalar product from
recoupling theory in loop quantum gravity," Phys. Rev. D 54(4), 2664-2690 (1996);
 S. Fittelli, L. Lehner, C. Rovelli, The complete spectrum of the area from recoupling
theory in loop quantum gravity," Class. Quant. Grav. 13, 2921-2932 (1996);
Abhay Ashtekar and Jerzy Lewandowski, Quantum Theory of Geometry I: Area operators,"
Class. Quant. Grav. 14, A55-A81 (1997).

\bibitem{ham}C. Rovelli and L. Smolin, The physical hamiltonian in nonperturbative 
quantum gravity,
Phys. Rev. Lett.72(1994)446; Spin Networks and Quantum Gravity, Phys. Rev.
D52(1995)5743-5759.

\bibitem{roumen-ham}R. Borissov, Graphical Evolution of Spin Network 
States, Phys.Rev. D55 (1997) 6099-6111, gr-qc/9606013.  
    
\bibitem{thomas-ham}T. Thiemann, Quantum Spin Dynamics (QSD) I \& II, 
Class.Quant.Grav. 15 (1998) 839-905, gr-qc/9606089,
gr-qc/9606090.

\bibitem{barbero}J. Barbero, Real Ashtekar variables for Lorentzian signature spacetime,"
Phys. Rev. D51 (1995) 5507.

\bibitem{mike-foam}M. P. Reisenberger. 
``Worldsheet formulations of gauge theories and gravity,"
in Proceedings of the 7th Marcel Grossman Meeting, ed. by R.
Jantzen and G. MacKeiser, World Scientific, 1996; gr-qc/9412035;
A lattice worldsheet sum for 4-d Euclidean general
relativity. gr-qc/9711052.

\bibitem{RR-foam}8] M. P. Reisenberger and C. Rovelli. Sum-over-surface form of loop quantum
gravity," gr-qc/9612035, Phys. Rev. D 56 (1997) 3490; Spacetime as a Feynman diagram:
the connection formulation. Class.Quant.Grav., 18:121{140, 2001;
Spin foams as Feynman diagrams, gr-qc/0002083.

\bibitem{BC}J. Barrett and L. Crane, ``Relativistic spin networks and 
quantum gravity'', J. Math. Phys.
39 (1998) 3296-3302, gr-qc/9709028.

\bibitem{baezfoam}J. Baez, Spin foam models, Class. Quant. Grav. 15 
(1998) 1827-1858, gr-qc/9709052;  An introduction to spin foam models of quantum gravity
and BF theory. Lect.Notes Phys., 543:25{94, 2000.


\bibitem{F-foam}Fotini Markopoulou, ``Dual formulation of spin network evolution'',
gr-qc/9704013.

\bibitem{tubes} Fotini Markopoulou, Lee Smolin, 
``Quantum geometry with intrinsic local causality'', Phys.Rev. D58 (1998) 
084032,  gr-qc/9712067.

\bibitem{foamreviews}J. Iwasaki, A reformulation of the Ponzano-Regge quantum gravity model
in terms of surfaces," gr-qc/9410010; A definition of the Ponzano-Regge
quantum gravity model in terms of surfaces," gr-qc/9505043, J. Math.
Phys. 36 (1995) 6288;
L. Freidel and K. Krasnov. Spin foam models and the classical action
principle. Adv.Theor.Math.Phys., 2:1183{1247, 1999; R. De Pietri, L. Freidel, 
K. Krasnov, and C. Rovelli. Barrett-Crane
model from a boulatov-ooguri field theory over a homogeneous space.
Nucl.Phys. B, 574:785{806, 2000.

\bibitem{kirill-radiate}Kirill Krasnov, ``Quantum Geometry and Thermal Radiation 
from Black Holes'', gr-qc/9710006, Class.Quant.Grav. 16 (1999) 563-578.

\bibitem{logcorrect}S. Das, Parthasarathi Majumdar, Rajat K. Bhaduri, `` 
General Logarithmic Corrections to Black Hole Entropy'', 
Class.Quant.Grav. 19 (2002) 2355-2368;  Saurya Das, hep-th/0207072,
``Leading Log Corrections to Bekenstein-Hawking Entropy''

\bibitem{fineBH}M. Barreira, M. Carfora, C. Rovelli, 
``Physics with nonperturbative quantum gravity: radiation from a quantum black hole''
gr-qc/9603064, Gen.Rel.Grav. 28 (1996) 1293-1299.

\bibitem{LQC}Martin Bojowald,''Isotropic Loop Quantum Cosmology'',
Class.Quant.Grav. 19 (2002) 2717-2742,   gr-qc/0202077; 
    ``Inflation from Quantum Geometry'', gr-qc/0206054;
    ``The Semiclassical Limit of Loop Quantum Cosmology'', 
gr-qc/0105113, Class.Quant.Grav. 18 (2001) L109-L116; 
`` Dynamical Initial Conditions in Quantum Cosmology'',
gr-qc/0104072,  Phys.Rev.Lett. 87 (2001) 121301.  

\bibitem{gangof5}A Ashtekar J Lewandowski D Marlof J Mourau 
T Thiemann:"Quantization of dieomorphism invariant theories of connections with local 
degrees of freedom", gr-qc/9504018, JMP 36 (1995) 519.

\bibitem{thomas}T. Thiemann, ``Quantum spin dynamics," gr-qc/9606089, gr-qc/9606090;
``Anomaly free formulation of non-perturbative 4-dimensional Lorentzian
quantum gravity," Phys. Lett. B380 (1996) 257.

\bibitem{thomas-thesis}Thomas Thiemann, 
``Introduction to Modern Canonical Quantum General Relativity'' gr-qc/0110034 

\bibitem{2+1}A. Ashtekar, V. Husain, C. Rovelli, J. Samuel and L. Smolin 2+1 quantum
gravity as a toy model for the 3+1 theory Class. and Quantum Grav. L185-
L193 (1989); L. Smolin, Loop representation
for quantum gravity in 2+1 dimensions,in the proceedings of the John's
Hopkins Conference on Knots, Tolopoly and Quantum Field Theory ed. L.
Lusanna (World Scientific,Singapore,1989) .

\bibitem{1+1}Viqar Husain, ``Observables for spacetimes with two Killing 
field symmetries'', gr-qc/9402019,Phys.Rev. D50 (1994) 6207-6216;
V. Husain and L. Smolin, ``Exactly Solvable Quantum Cosmologies From 
Two Killing Field spacetimes'', Nucl.Phys. B327 (1989) 205.


\bibitem{matter}R. Gambini, Phys. Lett. B 171 (1986) 251; P. J. Arias, C. Di Bartolo,
X. Fustero, R. Gambini and A. Trias, Int. J. Mod. Phys. A 7 (1991)
737.
[38] L. Smolin, Phys. Rev. D 49 (1994) 4028.

\bibitem{latticeloop}R. Gianvittorio, R. Gambini and A. Trias, Phys. Rev. D38 (1988) 702; C.
Rovelli and L. Smolin. Loop representation for lattice gauge theory,1990
Pittsburgh and Syracuse preprint; B. Bruegmann, Physical Review D 43
(1991) 566; J.M.A. Farrerons, Loop calculus for SU(3) on the lattice Phd.
thesis, Universitat Autonoma de Barcelona (1990); R. Loll A new quantum
representation for canonical gravity and SU(2) Yang-Mills theory, University
of Bonn preprint, BONN-HE-90-02 (1990)

\bibitem{super}T. Jacobson, New Variables for canonical supergravity, Class. Quant. 
Grav.5(1988)923-
935; D. Armand-Ugon, R. Gambini, O. Obregon, J. Pullin, Towards a loop representation
for quantum canonical supergravity, hep-th/9508036, Nucl. Phys. B460 (1996) 615; L.
F. Urrutia Towards a loop representation of connection theories defined over a super-lie
algebra, hep-th/9609010;  H. Kunitomo and T. Sano 
The Ashtekar formulation for canonical N=2 supergravity,
Prog. Theor. Phys. suppl. (1993) 31; Takashi Sano and J. Shiraishi, The Nonperturbative
Canonical Quantization of the N=1 Supergravity, Nucl. Phys. B410 (1993)
423, hep-th/9211104; The Ashtekar Formalism and WKB Wave Functions of N=1,2
Supergravities, hep-th/9211103; T. Kadoyoshi and S. Nojiri, N=3 and N=4 two form
supergravities, Mod. Phys. Lett. A12:1165-1174,1997, hep-th/9703149; 
K. Ezawa, Ashtekar's formulation for N=1, N=2 supergravities as constrained BF theories,
Prog. Theor. Phys.95:863-882, 1996, hep-th/9511047.

\bibitem{yi-super}Yi Ling, ``Introduction to supersymmetric spin networks'',
hep-th/0009020,  J.Math.Phys. 43 (2002) 154-169

\bibitem{higher} L. Freidel, K. Krasnov, R. Puzio,
`` BF Description of Higher-Dimensional Gravity Theories'',
hep-th/9901069, Adv.Theor.Math.Phys. 3 (1999) 1289-1324.

\bibitem{11d}Y. Ling and L. Smolin, Eleven 
dimensional supergravity as a constrained topological field
theory, hep-th/0003285, Nucl.Phys. B601 (2001) 191-208

\bibitem{stringsfrom} Lee Smolin, ``Strings as perturbations of evolving 
spin-networks'', Nucl.Phys.Proc.Suppl. 88 (2000) 103-113, 
hep-th/9801022.

\bibitem{MCS}L. Smolin, ``M theory as a matrix extension of Chern-Simons 
theory'',  hep-th/0002009, Nucl.Phys. B591 (2000) 227-242; 
``The cubic matrix model and a duality between strings and loops'', 
hep-th/0006137; ``The exceptional Jordan algebra and the matrix string'', 
hep-th/0104050.

\bibitem{louis-holo}L. Crane in, Categorical Physics, hep-th/9301061; 
Topological Field theory as the key
to quantum gravity, hep-th/9308126, in Knot theory and quantum gravity ed. J. Baez,
(Oxford University Press); Clocks and Categories, is quantum gravity algebraic? J.
Math. Phys. 36 (1995) 6180-6193, gr-qc/9504038.

\bibitem{carlo-relational}Carlo Rovelli, ``Relational Quantum Mechanics''
quant-ph/9609002,Int. J. of Theor. Phys. 35 (1996) 1637.

\bibitem{Fotini-QCH} Fotini Markopoulou,'' An insider's guide to quantum 
causal histories'',hep-th/9912137, Nucl.Phys.Proc.Suppl. 88 (2000) 308-313;
``Quantum causal histories'',hep-th/9904009, Class.Quant.Grav. 17 (2000) 2059-2072;
``The internal description of a causal set: What the universe looks like from the inside'',
Commun.Math.Phys. 211 (2000) 559-583, gr-qc/9811053.


\bibitem{pluralism}L. Smolin, The Bekenstein Bound, Topological Quantum 
Field Theory and Pluralistic
Quantum Field Theory, gr-qc/9508064.

\bibitem{weakholo}F. Markopoulou and L. Smolin, Holography in a 
quantum spacetime, hep-th/9910146.

\bibitem{weakstrong}L. Smolin, The strong 
and the weak holographic principles, hep-th/0003056.

\bibitem{dirac}P. A. M. Dirac, Lectures on Quantum Mechanics Belfer Graduate School
of Science Monographs, no. 2 (Yeshiva University Press, New York,1964).

\bibitem{stachel-bi}J. Stachel, ``Einstein's search for general 
covariance, 1912-15'' in {\it Einstein and the History of General 
Relativity} vol 1 of {\it Einstein Studies} eds. D. Howard and J. 
Stachel. (Birkhauser,Boston,1989).

\bibitem{threeroads}L. Smolin {\it Three Roads to Quantum Gravity}
(Weidenfeld and Nicolson and Basic Books, London and New York,2001)

\bibitem{basic-tft}M. Atiyah, Topological quantum field theory Publ. Math. IHES 68
(1989) 175; The Geometry and Physics of Knots, Lezion Lincee (Cambridge
University Press, Cambridge,1990); G. Segal, Conformal field
theory oxford preprint (1988).

\bibitem{5cs}R. Floreanini and R. Percacci, Phys. Lett. B224 (1989) 291-294;
B231:119-124,1989.
[35] V.V. Fock, N.A. Nekrasov, A.A. Rosly, K.G. Selivanov What we
think about the higher dimensional Chern-Simons theories (Moscow,
ITEP). ITEP-91-70, July 1991. 7pp. in Sakharov Conf.1991:465-472
(QC20:I475:1991)

\bibitem{highercs}M. Banados, M. Henneaux, C. Iannuzzo and C. M. Viallet, A note on
the gauge symmetries of pure Chern-Simons theory with p-form gauge
elds gr-qc/9703061;
Max Banados,Luis J. Garay and Marc Henneaux, Nucl.Phys.B476:611-635,1996,
hep-th/9605159; Phys.Rev.D53:593-596,1996, hep-th/9506187.

\bibitem{AL}J. Ambjorn, A. Dasgupta, J. Jurkiewiczcy and R. Loll, 
``A Lorentzian cure for Euclidean troubles'', hep-th/0201104;
J. Ambjorn and R. Loll, Nucl. Phys. B536 (1998) 407 [hep-th/9805108];
J. Ambjorn, J. Jurkiewicz and R. Loll,
Phys. Rev. Lett. 85 (2000) 924 [hepth/ 0002050]; Nucl. Phys. B610 (2001) 347
[hep-th/0105267]; R. Loll, Nucl. Phys. B (Proc. Suppl.) 94
(2001) 96 [hep-th/0011194]; J. Ambjorn, J. Jurkiewicz and R. Loll, Phys.
Rev. D64 (2001) 044011 [hep-th/0011276]; JHEP 09 (2001) 022 [hep-th/0106082].

\bibitem{fotini-rg}Fotini Markopoulou,''Coarse graining in spin foam models'',
gr-qc/0203036; ``An algebraic approach to coarse graining'',
hep-th/0006199.

\bibitem{problem1}L. Smolin, The classical limit and the form of 
the hamiltonian constraint in nonperturbative 
quantum gravity, gr-qc/9607034.

\bibitem{problem2}J. Lewandowski, D. Marolf, Loop constraints: a habitat and their algebra,"
gr-qc/9710016;  R. Gambini, J. Lewandowski, D. Marolf, J. Pullin, On the consistency of
the constraint algebra in spin-network quantum gravity," gr-qc/9710018.

\bibitem{mewheeler}L. Smolin, to appear in the procedings of the 
conference in honor of John Wheeler, Princeton 2002.

\bibitem{abhay-books}A. Ashtekar, 
New perspectives in canonical gravity (Bibliopolis, Naples,
1988); Lectures on non-perturbative canonical gravity, Advanced Series in Astrophysics
and Cosmology-Vol. 6 (World Scientific, Singapore, 1991).

\bibitem{AndyetdS}M. Spradlin, A. Strominger, A. Volovich1y, 
Les Houches Lectures on de Sitter Space, hep-th/0110007



\bibitem{BF}G. Horowitz,  ``Exactly Soluble Diffeomorphism Invariant 
Theories'', Commun.Math.Phys. 125 (1989) 417; V. Husain, ``Topological
Quantum Mechanics'',  Phys.Rev. D43 (1991) 1803.

\bibitem{JSS}T. Jacobson and L. Smolin,
Phys. Lett. B 196 (1987) 39; Class. and Quant. Grav. 5 (1988) 583;
J. Samuel, Pramana-J Phys. 28 (1987) L429.

\bibitem{Plebanski}J.F. Plebanski. On the separation of einsteinian substructures. J. Math.
Phys., 18:2511, 1977.

\bibitem{CDJ}R. Capovilla, J. Dell and T. Jacobson, Phys. Rev. Lett. 21, 
2325(1989); Class. Quant.
Grav. 8, 59(1991); R. Capovilla, J. Dell, T. Jacobson and L. Mason, 
Class. and Quant.
Grav. 8, 41(1991).

\bibitem{BGP}B. Bruegmann, R. Gambini and J. Pullin, Phys. Rev. Lett. 68 (1992) 431-
434;

\bibitem{banks1}T. Banks,  ``T C P, Quantum Gravity, The Cosmological Constant 
And All That'',  Nucl.Phys. B249 (1985) 332.

\bibitem{gravitons}A. Ashtekar, C. Rovelli and L. Smolin ``Gravitons and Loops'', Phys. Rev. D
44 (1991) 1740-1755; J. Iwasaki, C. Rovelli, ``Gravitons as embroidery on the weave," Int. J.
Mod. Phys. D 1 (1993) 533; ``Gravitons from loops: non-perturbative loop-space
quantum gravity contains the graviton-physics approximation," Class.
Quantum Grav. 11 (1994) 1653.

\bibitem{newcarlolee}C. Rovelli and L. Smolin, in preparation.

\bibitem{GP}Rodolfo Gambini, Jorge Pullin, ``Nonstandard optics from quantum 
spacetime'', Phys.Rev. D59 (1999) 124021, gr-qc/9809038;  

\bibitem{AMU}Jorge Alfaro, Hugo A. Morales-TŽcotl, Luis F. Urrutia,
``Loop quantum gravity and light propagation'',  Phys.Rev. D65 (2002) 
103509, hep-th/0108061  

\bibitem{gac-dsr}G. Amelino-Camelia, Nature 418 (2002) 34.

\bibitem{dsr2}N.R. Bruno, G. Amelino-Camelia, J. Kowalski-Glikman,
Phys.Lett.B522:133-138,2001; J. Kowalski-Glikman and 
S. Nowak, hep-th/0203040;S. Judes, gr-qc/0205067; 
M. Visser, gr-qc/0205093; S. Judes, M. Visser, gr-qc/0205067;
D. V. Ahluwalia and M. Kirchbach, qr-qc/0207004.

\bibitem{joaolee1}J. Magueijo and  L. Smolin,
Phys.Rev.Lett. (88) 190403, 2002.

\bibitem{joaolee2}J. Magueijo and  L. Smolin, gr-qc/0207

\bibitem{GH}G. W. Gibbons and S. W. Hawking, 
Cosmological Event Horizons, Thermodynamics,
and Particle Creation," Phys. Rev. D 15, 2738 (1977).

\bibitem{chopinthesis}L. N. Chang and C. Soo, 
Ashtekar's variables and the topological phase of quantum gravity, Proceedings of the XXth.
Conference on Differential Geometric Methods in Physics, (Baruch College, 
New York, 1991), edited by S. Catto and A.
Rocha (World Scientific, 1992); Phys. Rev. D46 (1992) 4257;
C. Soo and L. N. Chang, Int. J. Mod. Phys. D3 (1994) 529.

\bibitem{lattice-sn}J. Kogut and L. Susskind, Phys. Rev. D 11 (1975) 395;
W. Furmanski and A. Kowala, Nucl. Phys. B 291 (1987) 594.

\bibitem{carlo-subtle}N Grott C Rovelli,
``Moduli space structure of knots with intersections'', J. Math. Phys. 37 (1996) 
3014, gr-qc/9604010.

\bibitem{witten-cs}E. Witten Quantum Field theory and the Jones polynomial in the Proceedings
of the 1988 IAMP Congress, Swansea; Commun. Math. Phys. 121
(1989)351.

\bibitem{lou-bracket}Louis H. Kauffman, Knots and Physics, Series on Knots and 
Everything - Vol. 1 (World
Scientific, Singapore, 1991) pp. 125-130, 443-471.

\bibitem{mecs}L. Smolin,
``Invariants Of Links And Critical Points Of The Chern-Simons Path Integrals,''
Mod.\ Phys.\ Lett.\ A {\bf 4}, 1091 (1989).

\bibitem{review-cs}S. Axelrod and I. M. Singer,
Chern-Simons Perturbation Theory, Proceedings of the XXth. Conference on 
Differential
Geometric Methods in Physics, (Baruch College, New York, 1991) edited by S. Catto 
and A. Rocha, 
(World Scientific,
1992); S. Axelrod and I. M. Singer, J. Diff. Geo. 39 (1994) 173; 
D. Bar-Natan, Perturbative aspects of the Chern-Simons topological quantum 
field theory, Ph. D. 
Thesis, Princeton Univ.,
June 1991; L. Rozansky, Commun. Math. Phys. 171 (1995) 279; 
Witten's Invariant of 3-Dimensional Manifolds: Loop Expansion and Surgery Calculus hep-th/9401060.

\bibitem{louis-2d3d}L. Crane, Commun.
Math. Phys. 135 (1991) 615Phys. Lett. B 259 (1991) 243.

\bibitem{KL}Louis H. Kauffman and Sostenes L. Lins, Temperley-Lieb Recoupling 
Theory and Invariants
of 3-Manifolds, Annals of Mathematics Studies N. 134, (Princeton University
Press, Princeton, 1994), pp. 1-100.

\bibitem{sethme}S. Major and L. Smolin, Quantum 
deformation of quantum gravity, Nucl. Phys. B473,
267(1996), gr-qc/9512020;

\bibitem{sethroumenme}R. Borissov, S. Major and L. Smolin, The geometry of quantum
spin networks, Class. and Quant. Grav.12, 3183(1996), gr-qc/9512043.

\bibitem{bb}J. D. Bekenstein, Lett. Nuovo. Cim 11(1974).

\bibitem{verlinde}E. Verlinde, Fusion rules and modular transformations in 
2D conformal eld theory Nucl. Phys. B 300 (1988) 360.

\bibitem{inprogress}L. Smolin, in preparation.

\bibitem{Dyson}F. Dyson, to appear in the procedings of the 
conference in honor of John Wheeler, Princeton 2002.

\bibitem{BI} C.J. Isham, J.Butterfield, 
``Some Possible Roles for Topos Theory in Quantum Theory and Quantum 
Gravity'', 
gr-qc/9910005, Found.Phys. 30 (2000) 1707-1735;
``A topos perspective on the Kochen-Specker theorem: II. Conceptual Aspects, and Classical Analogues''
quant-ph/9808067, Int.J.Theor.Phys. 38 (1999) 827-859; 
``A topos perspective on the Kochen-Specker theorem: I. Quantum States 
as Generalized Valuations'', quant-ph/9803055;  C.J. Isham, 
``Topos Theory and Consistent Histories: The Internal Logic of the Set 
of all Consistent Sets'',
gr-qc/9607069, Int.J.Theor.Phys. 36 (1997) 785-814.

\bibitem{stume}Stuart Kauffman, Lee Smolin, ``Combinatorial dynamics in quantum gravity'',
hep-th/9809161,  Lect.Notes Phys. 541 (2000) 101-129.

\bibitem{qc-cs}R. Graham and R. Paternoga, Phys. Rev. D 54, 2589 (1996);
 D 54, 4805 (1996);  D 58 083501 (1998).

\bibitem{triad}R. Paternoga and R. Graham, Triad representation of the 
Chern-Simons state in quantum gravity, gr-qc/0003111

\bibitem{GP-lambda}R. Gambini and J. Pullin, Phys. Rev. Lett. 85
(2000) 5272, Class. Quant. Grav. 17 (2000) 4515.


\bibitem{GP-generate}Rodolfo Gambini, Jorge Griego, Jorge Pullin,
``Chern--Simons states in spin-network quantum gravity'', gr-qc/9703042,
Phys.Lett. B413 (1997) 260-266; 
 C. Di Bartolo, R. Gambini, J. Griego, J. Pullin, 
``Consistent canonical quantization of general relativity in the space of Vassiliev knot invariants'',
gr-qc/9909063, Phys.Rev.Lett. 84 (2000) 2314-2317; 
``Canonical quantum gravity in the Vassiliev invariants arena: I. Kinematical 
structure'', gr-qc/9911009,  Class.Quant.Grav. 17 (2000) 3211-3238.
} } } }

\end{thebibliography}
\end{document}